\documentclass[lettersize,journal]{IEEEtran}
\usepackage{amsmath,amsfonts}
\usepackage{algorithmic}
\usepackage{algorithm}
\usepackage{array}
\usepackage[caption=false,font=normalsize,labelfont=sf,textfont=sf]{subfig}
\usepackage{textcomp}
\usepackage{stfloats}
\usepackage{url}
\usepackage{verbatim}
\usepackage{graphicx}
\usepackage{booktabs}
\usepackage{multirow} 
\usepackage{cite}
\begin{document}

\title{PCDM: A Diffusion-Based Data Poisoning Attack\\ Against Federated Learning Systems}

\author{Wei Sun*, Yijun Chen*, Bo Gao{$\dagger$}, \IEEEmembership{Member, IEEE}, Ke Xiong, \IEEEmembership{Member, IEEE}, Yuwei Wang \\Pingyi Fan, \IEEEmembership{Senior Member, IEEE}, Khaled Ben Letaief, \IEEEmembership{Fellow, IEEE}

        \thanks{* Equal contribution. \textsuperscript{$\dagger$} Corresponding author. }

	\thanks{W. Sun, Y. Chen, B. Gao and K. Xiong are with the Engineering Research Center of Network Management Technology for High Speed Railway of Ministry of Education, School of Computer Science and Technology, and the Collaborative Innovation Center of Railway Traffic Safety, Beijing Jiaotong University, Beijing 100044, China. E-mail: \{21120398, 24120314, bogao, kxiong\}@bjtu.edu.cn.}
	\thanks{Y. Wang is with the Institute of Computing Technology, Chinese Academy of Sciences, Beijing 100190, China. E-mail: ywwang@ict.ac.cn.}
    \thanks{P. Fan is with the Beijing National Research Center for Information Science and Technology, and the Department of Electronic Engineering, Tsinghua University, Beijing 100084, China. E-mail: fpy@tsinghua.edu.cn.}
	\thanks{K. Letaief is with the Department of Electrical and Computer Engineering, Hong Kong University of Science and Technology, Hong Kong 999077, China. Email: eekhaled@ust.hk}}

\maketitle

\begin{abstract}
Federated learning (FL) is vulnerable to data poisoning attacks due to its distributed nature. Although recent GAN-based data poisoning methods have indicated the potential of using generative AI to generate seemingly legitimate poisoned data, the inherent consistency of GAN outputs can still reveal a sign of data poisoning. In this paper, we propose a diffusion-based data poisoning framework against FL systems, which leverages a Poisoning-Oriented Conditional Diffusion Model (PCDM) to enable fine-grained control over the local generation of poisoned data while ensuring both attack effectiveness and stealthiness. Our PCDM incorporates an adjustable poisoning vector within the global context to precisely control the generation of poisoned data, with theoretical guarantees on attack performance. Furthermore, it employs a novel jumping diffusion strategy for lightweight and efficient poisoned data generation. We conduct the most systematic and broad experimental evaluation for FL poisoning attacks against various defenses, including advanced Byzantine robust aggregation mechanisms, on four open datasets: MNIST, Fashion-MNIST, CIFAR-10, CIFAR-100, and a real-world wireless-specific dataset VRAI. Our results demonstrate that PCDM is less likely to exhibit statistical anomalies compared with the state-of-the-art methods while more effectively degrading global FL performance, which poses a significant risk to data security in FL.
\end{abstract}

% \begin{abstract}
% Federated learning (FL) enables collaborative model training across decentralized data sources without raw data exchange, but its distributed nature makes it vulnerable to data poisoning attacks. Although recent GAN-based data poisoning methods have indicated the potential of using generative AI to generate seemingly legitimate poisoned data, the inherent consistency of GAN outputs can still reveal a sign of data poisoning. In this paper, we propose a diffusion-based data poisoning framework against FL systems, which leverages a Poisoning-Oriented Conditional Diffusion Model (PCDM) to enable fine-grained control over the local generation of poisoned data while ensuring both attack effectiveness and stealthiness. Our PCDM incorporates an adjustable poisoning vector within the global context to precisely control the generation of poisoned data, with theoretical guarantees on attack performance. Furthermore, it employs a novel jumping diffusion strategy for lightweight and efficient poisoned data generation. Through the most systematic and broad experimental evaluation for FL poisoning attacks and defenses to date on four open datasets: MNIST, Fashion-MNIST, CIFAR-10 and CIFAR-100, we demonstrate that PCDM is less likely to exhibit statistical anomalies compared with the state-of-the-art methods, while more effectively degrading global FL performance, posing a significant risk to data security in FL and thus also provides an important research direction for future data protection in FL. 

% \end{abstract}

\begin{IEEEkeywords}
Data Poisoning, Security and Privacy, Generative Adversarial Networks, Federated Learning.
\end{IEEEkeywords}

\section{Introduction}
\label{submission}

\IEEEPARstart{W}{ith} the explosive growth of Internet of Things (IoT) and mobile devices, massive amounts of data are being generated at the wireless network edge. However, these data are typically fragmented across distributed devices, forming isolated data silos, and often contain privacy-sensitive information that restricts centralized collection. Federated Learning (FL) has emerged as a paradigm shift to unleash the potential of such distributed data. By coordinating collaborative training directly on wireless edge devices without exchanging raw data, FL significantly mitigates the privacy risks and reduces the burden on limited communication resources \cite{yazdinejad2022block}. 

While FL protects privacy by keeping data local, this opacity introduces significant security risks, especially in large-scale wireless systems. Due to the ubiquitous connectivity of 5G/6G networks, a vast number of heterogeneous edge devices can access the model training process. This uncontrolled participation implies that the server communicates with potentially compromised or malicious clients without access to their ground-truth data. As a result, FL is inherently vulnerable to data poisoning attacks initiated by these internal adversaries who exploit the lack of server supervision \cite{rodriguez2023survey,wei2024data}.

Data poisoning attacks maliciously tamper with local datasets, aiming to reduce the global FL performance. Theoretically, in the absence of the server's awareness, such tampering of local datasets by internal attackers (malicious clients) can degrade the performance of the global model, even if only a few malicious clients exist. There are two major types of data poisoning attacks in FL \cite{wei2023demystifying}: Targeted Data Poisoning Attacks (TDPA) and Untargeted Data Poisoning Attacks (UDPA). TDPA focuses on causing the global model to misclassify specific samples or categories, with the backdoor attack \cite{bagdasaryan2020backdoor} being the most representative TDPA in FL. In contrast, UDPA aims to degrade the overall performance of the global model. For example, label flipping attacks  \cite{tolpegin2020data} can be used to randomly flip the labels of training data, thereby corrupting the quality of training data and training outcomes. To explore more aggressive data poisoning attack methods and set a more threatening target for FL's security research, this paper investigates UDPA.

In the context of UDPA, both attack effectiveness and stealthiness are critical. However, existing attack methods often focus only on effectiveness while neglecting stealthiness \cite{li2024threats}. Traditional attack methods, such as label flipping attacks, typically cause significant changes in the distribution of local labels, making them easily detectable by existing defenses \cite{shen2016auror, upreti2022defending,chen2024exploring}. Although the recent Generative Adversarial Network (GAN)-based data poisoning, such as VagueGAN \cite{10287523}, have demonstrated the potential of using generative AI to generate seemingly legitimate poisoned data, the inherent consistency of GAN outputs can still reveal signs of data poisoning \cite{sun2024gan}. Current attack strategies fail to simultaneously achieve high impact and stealth. However, emerging Generative AI techniques, particularly GANs, have introduced a potent new threat vector capable of overcoming this limitation.

Given the limitations of existing attack methods in terms of stealthiness, in this paper, we propose to leverage diffusion models to generate inconsistent, seemingly legitimate poisoned data that is hard to detect. As an emerging generative AI technique distinct from GAN models, diffusion models generate data through a process of gradually adding noise and then denoising in reverse, allowing for better fine-grained control over the generated data \cite{cao2024survey}. Due to the stochastic nature of the generation process, diffusion models can be an ideal candidate for data poisoning attacks with enhanced stealthiness. 

The main contributions of this paper are as follows: 
\begin{itemize} 

	\item We propose an attack framework based on the Poisoning-Oriented Conditional Diffusion Model (PCDM), a diffusion model specifically designed for data poisoning attacks on FL systems. It generates poisoned data that can effectively and stealthily compromise a global model.

    \item We design a controllable and lightweight data poisoning mechanism for PCDM, specifically optimized for resource-limited wireless nodes. By incorporating an adjustable poisoning vector and a jumping diffusion strategy, this method enables precise poisoned data generation without burdening the edge devices.

	\item We develop guidelines for taking full advantage of PCDM to effectively achieve a balanced trade-off between attack effectiveness and stealthiness.
    \item We conduct extensive experiments on \textbf{five} datasets: MNIST, Fashion-MNIST, CIFAR-10, CIFAR-100, and a real-world wireless-specific dataset VRAI, evaluating \textbf{eleven} defense methods (including \textbf{three} robust aggregation mechanisms) against seven poisoning attacks. To our knowledge, this is the most comprehensive study to date. Results show that PCDM more effectively degrades FL performance with fewer anomalies and largely evades existing defenses.
	%\item We conduct extensive experiments on four popular datasets: MNIST, Fashion-MNIST, CIFAR-10, and CIFAR-100, evaluating eight defense methods against seven poisoning attacks. To our knowledge, this is the most comprehensive study to date. Results show that PCDM more effectively degrades FL performance with fewer anomalies and largely evades existing defenses.
\end{itemize}

This paper is organized as follows: Section II presents a review of related work on data poisoning attacks and defense strategies in FL. Section III outlines the attack objectives and system assumptions, and introduces the foundational concepts of denoising diffusion probabilistic models. In Section IV, we propose the PCDM, along with a rigorous theoretical analysis of its attack effectiveness. Section V evaluates the performance of PCDM through extensive experiments, including comparative and ablation studies. Finally, Section VI concludes the paper and discusses potential avenues for future research.

\section{Related Work}

FL is a promising research area \cite{li2020review,liu2024vertical,wen2023survey,liu2024recent}. However, potential data poisoning attacks pose a significant threat to FL \cite{nowroozi2025federated,kasyap2024beyond,wan2024data,zhang2024visualizing}. In data poisoning attacks \cite{wu2024challenges}, the attacker exploits the distributed nature of FL to influence the training quality of the global model by constructing poisoned data on the client. 

In FL data poisoning, attack methods can be classified into two main categories: direct methods and AI-driven poisoning methods. On the one hand, direct poisoning attack methods introduce malicious updates into the global model by modifying training data directly. One common method is label flipping attack \cite{tolpegin2020data}, which constructs poisoned data by disrupting the correspondence between samples and labels. In \cite{10287523}, the authors construct poisoned data by superimposing various types of noise onto real data. On the other hand, AI-driven poisoning methods leverage advanced AI techniques to create poisoned data that is statistically indistinguishable from original data. Hyperdimensional data poisoning attack (HDPA) \cite{KASYAP2024122210} projects real data into a hyperdimensional computing space and adds perturbations to generate poisoned data. PoisonGAN \cite{zhang2020poisongan} is a poisoned data augmentation method to improve the effectiveness of data poisoning attacks, such as the label flipping attacks. However, the poisoned models trained on these poisoned data usually deviate significantly from the benign models, leading to poor stealthiness of the attacks. Recently, VagueGAN \cite{10287523} turns to reversely leverage the power of GAN to generate ``vague data" for more effective and stealthy data poisoning attacks, but the inherent consistency of GAN outputs makes it still detectable according to model consistency anomalies \cite{borji2022pros,wang2023gan,zhang2024fltracer}. In summary, existing attack methods generally fail to succeed due to limited stealthiness.

Despite the varying implementations, existing attack methods usually share a common limitation: they inevitably introduce detectable anomalies in model updates. The deviations in gradient direction, magnitude, or distribution caused by poisoned data typically create significant discrepancies between poisoned and benign models. To defend against such data poisoning attacks, various anomaly detection techniques have been proposed. PCA-based methods \cite{tolpegin2020data} reduce the dimensionality of local model updates and detect outliers as poisoned models. UMAP \cite{upreti2022defending} provides a nonlinear alternative to PCA for dimensionality reduction, using cosine and Euclidean distance metrics to identify anomalous updates. CONTRA \cite{awan2021contra} computes cosine similarity among all clients and flags those with significantly deviating gradient directions. DnC \cite{shejwalkar2021manipulating} projects gradients onto principal components and removes clients with extreme projections. Defenses based on K-Means\cite{li2022robust, onsu2023cope} cluster gradients to isolate outliers after PCA. FedDMC \cite{mu2024feddmc} introduces a binary tree-based noise clustering approach after PCA to robustly detect poisoned models. LoMar\cite{li2021lomar} uses kernel density estimation to evaluate the malicious degree of each client's model update. MCD \cite{sun2024gan} periodically monitors the distribution of local model parameters to detect anomalous behaviors, including those from GAN-based poisoning. 

In addition to these detection-oriented algorithms, recent studies have expanded into adaptive federated defenses and adversarially robust aggregation. Early influential approaches such as Multi-Krum \cite{blanchard2017machine} rely on global distance metrics, screening out statistical outliers by selecting the subset of local updates with the smallest Euclidean distances to their neighbors. To move beyond purely distance-based statistics, SignGuard \cite{xu2022byzantine} combines direction-based clustering with magnitude-based constraints, leveraging gradient signal statistics to filter malicious model updates. More recently, LASA \cite{xu2025achieving} proposes a finer-grained defense through layer-adaptive aggregation. It incorporates pre-aggregation sparsification to reduce the attack surface and employs a layer-wise filter based on magnitude and direction purity to robustly identify benign parameters in non-IID environments. Additionally, substantial progress has been made in securing FL within specialized domains. Works focusing on healthcare and IoT \cite{yazdinejad2023ap2fl,yazdinejad2025breaking,yazdinejad2024robust,yazdinejad2025advanced,yazdinejad2025explainable,yazdinejad2024hybrid} have introduced novel mechanisms that yield excellent robustness outcomes in these practical settings.

However, since most of these defenses and frameworks rely on detecting statistical irregularities or structural anomalies in model updates, they become ineffective if the poisoning attack can stealthily manipulate poisoned updates to mimic benign behaviors. As summarized in Table~\ref{tab:attack_comparison}, existing attack methods generally suffer from limited stealthiness due to detectable artifacts such as magnitude deviations or distribution shifts. To bridge this gap, we propose PCDM, which generates highly covert poisoned data that bypasses these state-of-the-art detection mechanisms.

\begin{table}[t]
\caption{Comparison of limitations and stealthiness among different poisoning attacks.}
\label{tab:attack_comparison}
\centering
\resizebox{\columnwidth}{!}{
\begin{tabular}{l l c}
\toprule
\textbf{Method} & \textbf{Primary Limitation } & \textbf{Stealthiness} \\
\midrule
Label Flipping \cite{tolpegin2020data} & Label-sample mismatch & Low \\
Noise Injection \cite{10287523} & Statistical outliers in gradients & Low \\
PoisonGAN \cite{zhang2020poisongan} & Distribution shifts & Low \\
VagueGAN \cite{10287523} & Model consistency anomalies & Medium \\
\midrule
\textbf{PCDM (Ours)} & \textbf{Hard to distinguish from benign} & \textbf{High} \\
\bottomrule
\end{tabular}
}
\vskip -0.2in
\end{table}

\section{OBJECTIVES AND ASSUMPTIONS}

\subsection{System Model}

\begin{figure}[ht]
	\vskip -0.15in  % 控制图形上方的间距
	\begin{center}
		\centerline{\includegraphics[width=\columnwidth]{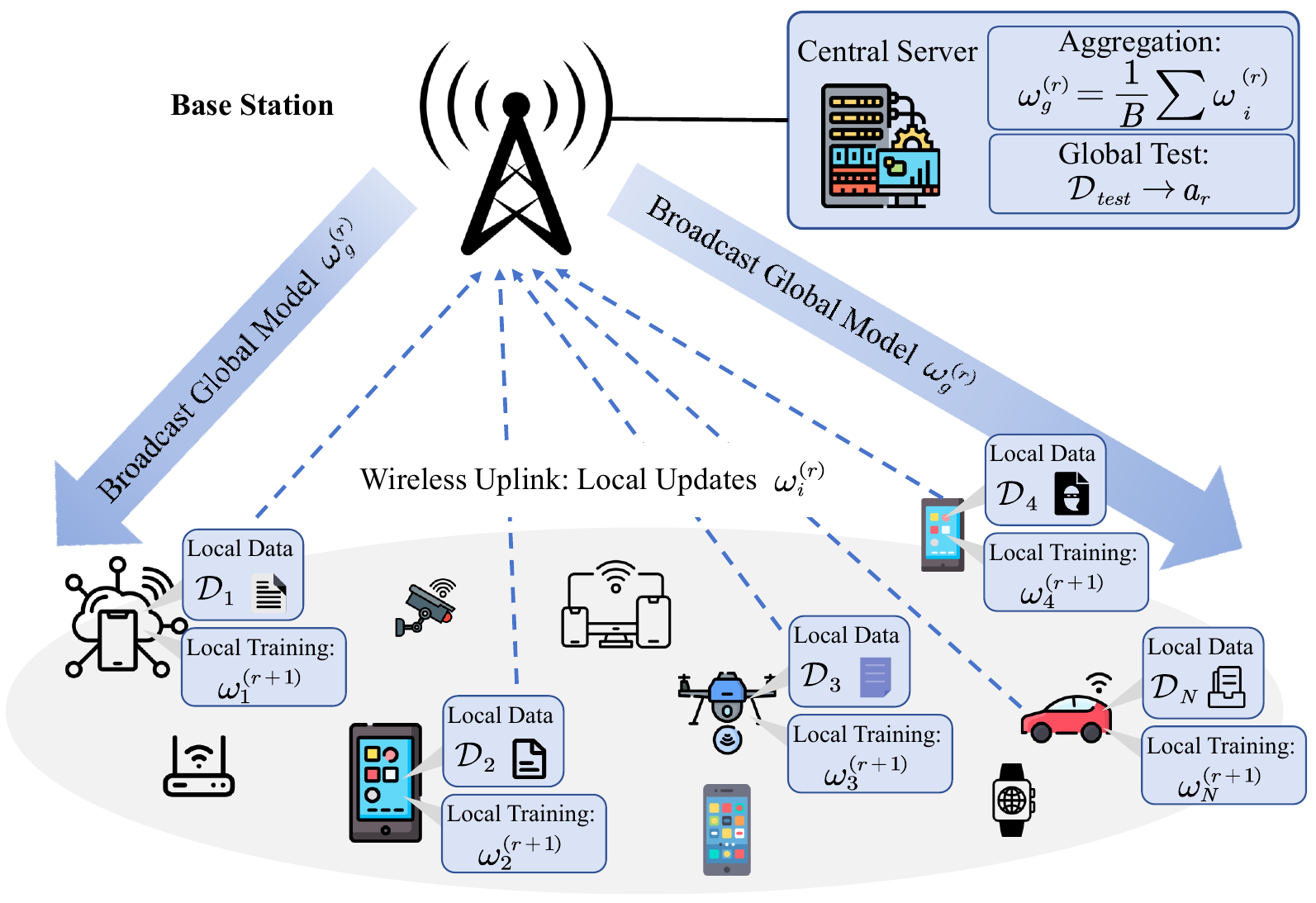}}
		\vskip -0.1in 
		\caption{Wireless federated learning system model.}
		\label{Ststem Model}
	\end{center}
	\vskip -0.3in  % 控制图形下方的间距
\end{figure}

In this research, we consider a wireless network-supported FL system comprising a central server equipped at the Base Station (BS) and $N$ distributed wireless clients, denoted as $\{c_1, c_2, \dots, c_N\}$. These clients utilize wireless uplinks to interact with the server. An overview of the considered system architecture is illustrated in Fig.~\ref{Ststem Model}. Each client $c_i$ possesses a local dataset $\mathcal{D}_i=\{(\mathbf{x}_k, y_k)_i \mid k=1, 2, \dots, K_i\}$ containing $K_i$ data samples. The training process in FL is synchronized across all devices, and the FL task trained on the central server and each client is identical, denoted as $F$. 

To enhance system efficiency and minimize communication overhead, in each round $r = 1, 2, \dots, R$, the server randomly selects $B<N$ clients from a total of $N$ clients, forming a subset denoted as $\mathcal{C}_r^B$. The server then applies the Federated Averaging (FedAvg) algorithm \cite{mcmahan2017communication} to compute the global model. Specifically, in the $r$-th round, the global model $\omega_g^{(r)}$ is calculated as:

%\vskip -0.1in
\begin{equation}
	\omega _{g}^{\left( r \right)}=\frac{1}{B}\sum_{c_i \in \mathcal{C}_r^B}{\omega}_{i}^{( r)}.
\end{equation}
% %\vskip -0.1in

The server then sends the global model $\omega _{g}^{\left( r \right)}$ to all the clients, which then train new local models based on $\omega _{g}^{\left( r \right)}$:
% %\vskip -0.2in
\begin{equation}
	\omega _{i}^{\left( r+1 \right)}=\omega_{g}^{\left( r \right)}-\eta \cdot \nabla F\left( \omega _{g}^{\left( r \right)},\mathcal{D}_i \right) ,
\end{equation}
where $\eta$, $F$, $\nabla F$ represent learning rate, learning task, and training gradient. Meanwhile, the server use the test dataset $\mathcal{D}_{test}$ to evaluate the global model accuracy $a_r$:

\begin{equation}
	a_r=F\left( \omega _{g}^{\left( r \right)},\mathcal{D}_{test} \right). 
\end{equation}

The objective of the server is to maximize the test accuracy of the global model $a_r$ during federated training, i.e., $\max _{\omega _{g}^{\left( r \right)}}F\left( \omega _{g}^{\left( r \right)},\mathcal{D}_{test} \right) $.

\subsection{Attack Model}

\begin{figure}[ht]
	\vskip -0.1in  % 控制图形上方的间距
	\begin{center}
		\centerline{\includegraphics[width=\columnwidth]{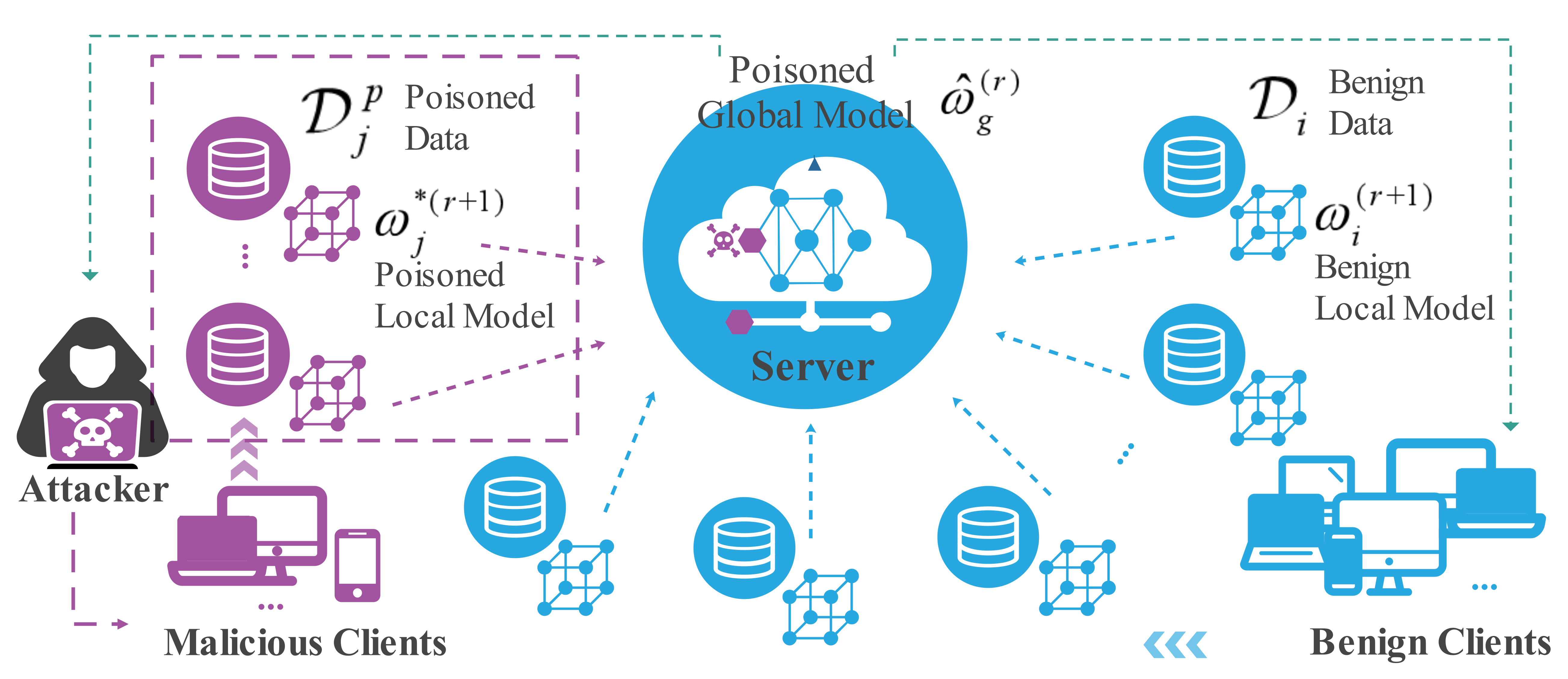}}
		\vskip -0.1in 
		\caption{The overall attack model.}
		\label{Attack Model}
	\end{center}
	\vskip -0.3in  % 控制图形下方的间距
\end{figure}

As depicted in Fig.~\ref{Attack Model}, 
due to the open and distributed nature of wireless networks, edge devices are susceptible to physical capture or software subversion. We consider a scenario where a subset of the participating wireless clients is compromised by an adversary, acting as malicious internal nodes. In the following, we characterize the attack model by detailing the adversary's targets, capabilities regarding network knowledge, and poisoning strategies deployed over the wireless links.

\textbf{Attacker’s targets}: We assume the presence of at least one attacker with two primary goals. Firstly, the attacker aims to degrade the overall performance of the global model by misclassifying samples. Secondly, it is equally important to hide the attacks from the server's detection.  

\textbf{Attacker’s capabilities}: The attacker can take control of or impersonate one or multiple benign clients. To emphasize the effectiveness of the data poisoning attack strategy, we assume that the controlled (malicious) clients are self-contained and do not involve the collaboration with external entities or resources. The local training process cannot be impacted by malicious clients, which can only manipulate their local datasets, i.e., through a black-box attack.

\textbf{Attacker’s approaches}: The attacker launches a UDPA by taking control of one or more malicious clients and using a generative model to locally construct poisoned dataset. The malicious clients then use the poisoned dataset to train poisoned local models and upload them to the server. Unless the server detects anomalies, the poisoned local models will compromise the global model after federated aggregation. 

% \begin{figure}[ht]
% 	%\vskip -0.15in  % 控制图形上方的间距
% 	\begin{center}
% 		\centerline{\includegraphics[width=\columnwidth]{figure1}}
% 		%\vskip -0.1in 
% 		\caption{The overall attack model.}
% 		\label{Attack Model}
% 	\end{center}
% 	\vskip -0.3in  % 控制图形下方的间距
% \end{figure}

\subsection{Denoising Diffusion Probabilistic Model}

\begin{figure}[ht]
\vskip -0.15in
	\begin{center}
		\centerline{\includegraphics[width=\columnwidth]{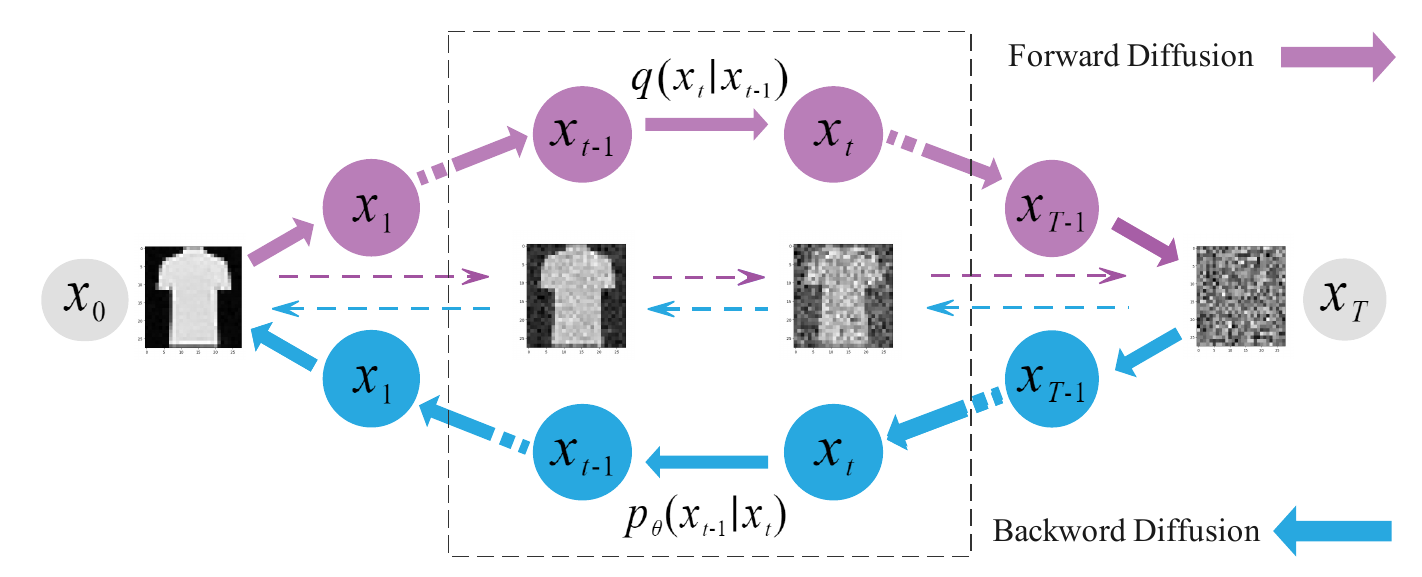}}
		\vskip -0.1in
		\caption{Denoising Diffusion Probabilistic Model (DDPM).}
		\label{fig:diff}
	\end{center}
	\vskip -0.3in
\end{figure}

Given the state-of-the-art performance of the Denoising Diffusion Probabilistic Model (DDPM) in the generative domain and its fine-grained control over generated samples \cite{regenwetter2022deep,bengesi2024advancements}, we aim to investigate the potential of malicious clients leveraging DDPM for poisoned data generation. As shown in Fig.~\ref{fig:diff}, DDPM is a deep generative model that uses variational inference to train a parameterized Markov chain. It has a forward diffusion process and a backward inverse diffusion process, with each process consisting of a limited number of time steps. The forward process is parameterless, while the backward process requires a training algorithm.

In the attack model, malicious clients train a DDPM using their local data, enabling them to generate poisoned data by denoising pure noise samples. Specifically, the process involves two stages: training and generation. The \textbf{training stage} consists of a forward process and a backward process:

Forward Diffusion Process: Suppose $\mathbf{x}_{0}\sim p(\mathbf{x})$ represents a sample from the distribution of a local dataset, and $t=1,2,..,T$ is a time variable, where $T$ is the maximum discrete time step during the diffusion process. The forward diffusion process gradually adds slight Gaussian noise $\epsilon \sim \mathcal{N}(0,\mathbf{I})$ to the original data $\mathbf{x}_{0}$, generating a series of noisy samples $[\mathbf{x} _ { 1 } , \mathbf{x} _ { 2 } ,\cdots ,\mathbf{x} _ { T-1 } , \mathbf{x} _ { T }]$ over $t = \{1 , 2 , \cdots ,T-1 ,T\}$ time steps. The variance sequence $\{ \beta _ { t } \in ( 0 , 1 ) \} _ { t = 1 } ^ { T }$ controls the stepsizes. The transition probability from $t-1$ to $t$ is denoted as $q ( \mathbf{x} _ { t } | \mathbf{x} _ { t - 1 } )$. Then the transition probability from the original data to the last step of the forward diffusion process is:
\vskip -0.1in
\begin{equation}
    q(\mathbf{x}_{1:T}|\mathbf{x}_{0})= \prod _{t=1}^{T}q(\mathbf{x}_{t}|\mathbf{x}_{t-1}),
\end{equation}
\vskip -0.1in
\noindent where,
\vskip -0.1in
\begin{equation}
	q(\mathbf{x}_{t}|\mathbf{x}_{t-1})=\mathcal{N}(\mathbf{x}_{t}; \sqrt{1- \beta _{t}}\mathbf{x}_{t-1}, \beta _{t} \mathbf{I}).
\end{equation}
\vskip -0.05in

As $t$ increases, the distinguishable features of data sample $\mathbf{x}_{0}$ gradually diminish. When $T \rightarrow \infty$,  $\mathbf{x}_{T}$ is equivalent to an isotropic Gaussian distribution. Then, use the reparameterization trick to sample $\mathbf{x}_{t}$ at any time step $t$. Let $\alpha _ { t } = 1 - \beta _ { t }$ and $\bar{\alpha} _ { t } = \prod _ { i = 1 } ^ { T } \alpha _ { i }$, so:
\vskip -0.1in
\begin{equation}\label{eq:6}
	q(\mathbf{x}_{t}|\mathbf{x}_{0})=\mathcal{N}(\mathbf{x}_{t}; \sqrt{\bar{\alpha}_{t}}\mathbf{x}_{0},(1- \bar{\alpha}_{t})\mathbf{I}).
\end{equation}
\vskip -0.05in
Backward Diffusion Process: The backward diffusion process is a parameterized Markov process, and the parameters of the transition probabilities are learned at each time step. It is impossible to estimate $q(\mathbf{x}_{t-1}|\mathbf{x}_{t})$ during the backward process without using the entire dataset. A model $p_{\theta}$ is needed to approximate the conditional probabilities for the backward diffusion process. Here let the time step $t = \{T,T-1,...,2,1\}$. The inverse diffusion process begins with noisy data at time step $t=T$, represented by the data distribution $p_{\theta}(\mathbf{x}_{T})=\mathcal{N}(\mathbf{x}_{T};0,\mathbf{I})$. The transition probability for the backward diffusion process is:
\vskip -0.1in
\begin{equation}
	p_{\theta}(\mathbf{x}_{0:T})=p(\mathbf{x}_{T})\prod _{t=1}^{T}p_{\theta}(\mathbf{x}_{t-1}|\mathbf{x}_{t}),
\end{equation}
\vskip -0.1in
\noindent where,
\vskip -0.2in
\begin{equation}
	p_{\theta}(\mathbf{x}_{t-1}|\mathbf{x}_{t})=\mathcal{N}(\mathbf{x}_{t-1}; \mu _{\theta}(\mathbf{x}_{t},t), \varSigma _{\theta}(\mathbf{x}_{t},t)).
\end{equation}
\vskip -0.05in
The \textbf{generation stage} begins by sampling the final output from the learned reverse process, starting with an initial sample $\mathbf{x}_{T}$ drawn from a Gaussian prior, $\mathbf{x}_{T} \sim \mathcal{N}(0, \mathbf{I})$. In this reverse process, the noisy sample $\mathbf{x}_{T}$ is progressively denoised into a poisoned sample $\mathbf{x}_{0}^p$ by sequentially generating intermediate samples $\mathbf{x}_{T-1}, \mathbf{x}_{T-2}, \dots, \mathbf{x}_{1}$ through iterative applications of the learned transition probabilities $p_{\theta}(\mathbf{x}_{t-1}|\mathbf{x}_{t})$. Both the mean $\mu_{\theta}(\mathbf{x}_{t},t)$ and variance $\Sigma_{\theta}(\mathbf{x}_{t},t)$ are parameterized by a neural network, typically a U-Net \cite{ho2020denoising}, which is optimized during training stage to approximate the true reverse process. The final output $\mathbf{x}_{0}^p$, representing the generated poisoned data, preserves the structural similarity of the original data while incorporating malicious modifications introduced during the training of the DDPM on the attacker's local dataset.

\section{Diffusion-based Data Poisoning Attack}

In general, the effectiveness of a poisoned model is inversely correlated with its stealthiness: more effective poisoned models usually exhibit greater divergence from benign models, thereby indicating lower attack  stealthiness, and vice versa. As a result, achieving both effective and stealthy data poisoning attacks remains challenging. For example, the recently proposed VagueGAN, which emphasizes improving attack stealthiness, has been shown to suffer from model consistency anomalies due to the inherent limitations of GAN models\cite{sun2024gan}. To address these limitations, we propose a novel diffusion-based data poisoning attack method, named Poisoning-Oriented Conditional Diffusion Model (PCDM). This method provides several notable advantages in the realm of data poisoning attacks: (1) PCDM retains the core characteristics of the original data while keeping poisoned data inconsistent and diversified, thereby ensuring the attack’s stealthiness. (2) While preserving these essential features, PCDM generates highly effective poisoned features, thereby enhancing the attack’s overall impact. (3) Through its configurable hyperparameters, PCDM allows precise modulation of the poisoning intensity to satisfy conflicting stealth and utility requirements. (4) PCDM adopts a lightweight diffusion strategy, significantly reducing resource and time consumption.

\subsection{Poisoning-Oriented Conditional Diffusion Model}

\begin{figure}[ht]
	%\vskip -0.1in
	\begin{center}
		\centerline{\includegraphics[width=\columnwidth]{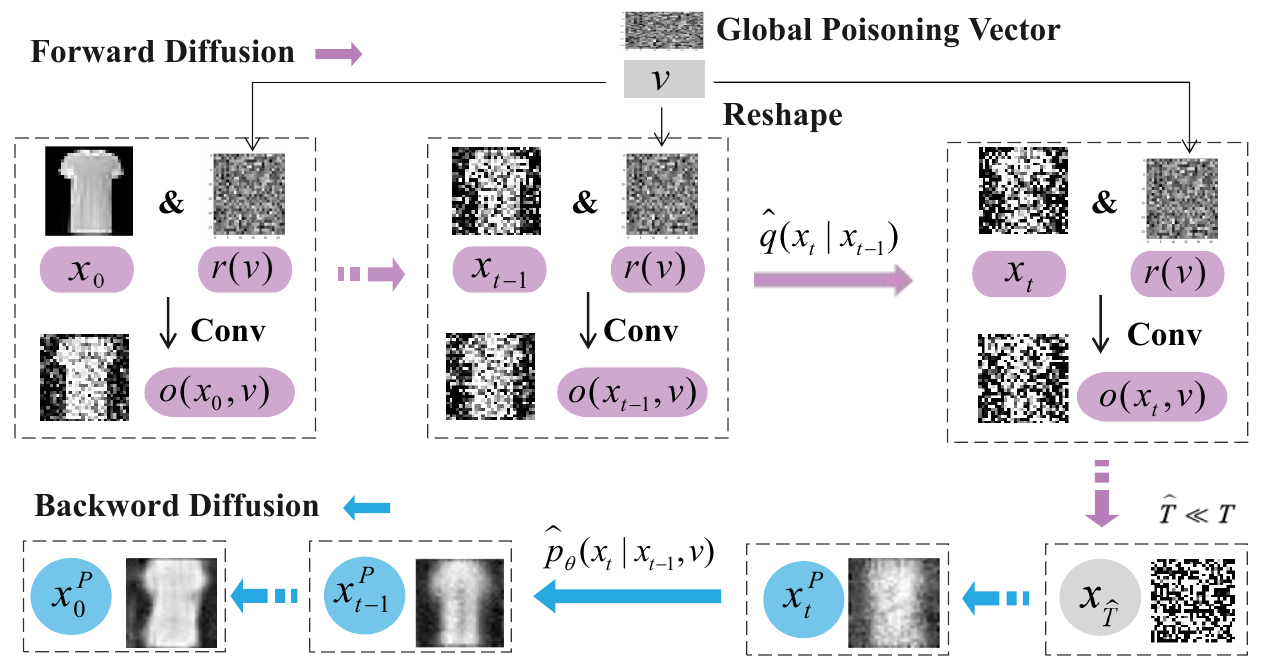}}
		\vskip -0.1in
		\caption{Workflow of the PCDM approach.}
		\label{fig:PCDM}
	\end{center}
	\vskip -0.4in
\end{figure}

Leveraging the superior capabilities of DDPM \cite{ho2020denoising} in the generative AI domain and its feature of fine-grained control over generated samples \cite{croitoru2023diffusion,yang2023diffusion}, DDPM shows significant potential for effectively carrying out data poisoning attacks. Specifically, DDPM uses a forward diffusion process that incrementally introduces noise to the data, and a reverse diffusion process, parameterized by a neural network, that iteratively removes noise and reconstructs the data. 

The significant training costs, prolonged data generation time, and substantial hardware requirements of DDPM pose challenges for its application in attacks. To address these issues, we developed PCDM, a model based on DDPM, to effectively generate poisoned data. As shown in Fig.~\ref{fig:PCDM}, our PCDM integrates a poisoning vector into the global context, which regulates the generation of poisoned features. Furthermore, PCDM employs an optimized jumping diffusion strategy, enabling the efficient generation of high-quality poisoned data with minimal resources, while substantially reducing computational costs compared to DDPM.

%\vskip 0.05in
\subsubsection{Poisoning Global Context} 

In PCDM, a poisoning vector $\mathbf{v}$ is introduced into the global context to guide the generation of poisoned data. This vector, a compact noise representation, is specifically designed to perturb the feature distribution, while the global context encodes high-level semantic information that informs and directs the overall generation process.

Specifically, at each time step $t$,  $\mathbf{v}$ is combined with the previous time step's output $\mathbf{x}_{t-1}$ to serve as the condition for computing the current output $\mathbf{x}_{t}$, thereby forming the context for the current time step:
\vskip -0.1in
\begin{equation}
	\mathbf{x}_t = f(\mathbf{x}_{t-1}, \mathbf{o}(\mathbf{x}_{t-1}, \mathbf{v})).
\end{equation}
\vskip -0.05in

In $\mathbf{o}(\cdot)$, the poisoning vector $\mathbf{v}$ is incorporated into the context computation. $f$ represents a function or transformation that governs the update of features at each time step $t$ in the diffusion model, $\mathbf{x}_{t-1}$ is the feature at time step $t-1$, $\mathbf{o}(\mathbf{x}_{t-1})$ is the context vector based on $\mathbf{x}_{t-1}$, and $\mathbf{v}$ is the poisoning vector injected into the model, as a global perturbation. In $\mathbf{o}(\cdot)$, the poisoning vector $\mathbf{v}$ is incorporated into the context computation by adding it to the features $\mathbf{x}_{t-1} $:
\vskip -0.1in
\begin{equation}
	\mathbf{o}(\mathbf{x}_{t-1},\mathbf{v}) = \text{Conv}(\mathbf{x}_{t-1} + r(\mathbf{v})).
\end{equation}
\vskip -0.05in
${Conv}$ denotes a convolutional layer that processes the combination of the feature $\mathbf{x}_t$ and the poisoning vector $\mathbf{v}$. Before the combination, $r(\cdot)$ reshapes $\mathbf{v}$ to align with the dimensions of $\mathbf{x}_t$. This reshaped vector, $r(\mathbf{v})$, introduces random perturbations that modify the feature extraction process.

Incorporating the poisoning vector introduces randomness into the generation process. Mathematically, the features of the poisoned data are given by:
\vskip -0.1in
\begin{equation}
	\mathbf{x}_t^{\text{poison}} = \mathbf{x}_t + \sigma_v \cdot \mathbf{v} + \mu_v,
\end{equation} 
where $\sigma_v$ and $\mu_v$ control the scale and shift of the poisoning vector, respectively. The intensity and direction of poisoning features in the generated poisoned data can be effectively controlled by adjusting the standard deviation $\sigma_v$ and the mean $\mu_v$ of the noise in the poisoning vector $\mathbf{v}$. This effect is quantified by interpreting the output distribution of the poisoning process as a noise function, where the conditional generation output $\mathbf{x}_t^p$ with respect to $\mathbf{v}$ is given by:
\vskip -0.15in
\begin{equation}
	p(\mathbf{x}_t^p | \mathbf{x}_t, \mathbf{v}) \propto \exp\left( - \frac{1}{2\sigma_v^2} \|\mathbf{x}_t^p - (\mathbf{x}_t + \mu_v + \mathbf{v})\|^2 \right).
\end{equation} 
\vskip -0.05in

This indicates that the poisoning vector $\mathbf{v}$ perturbs the distribution of the poisoned output, with the poisoning intensity $\sigma_v$ and mean $\mu_v$ controlling the degree of poisoning feature.

%\vskip 0.05in
\subsubsection{Properties of the Poisoning Vector}

In this subsubsection, we examine the properties of the poisoning vector $\mathbf{v}$ and perform a theoretical analysis of the characteristics of the poisoned data generated by PCDM.

\textbf{Property 1: Invariance of Forward Diffusion}\label{1111}

The inclusion of the poisoning vector $\mathbf{v}$ does not affect the forward diffusion process. This fundamental property ensures that the statistical nature of the forward dynamics, which meticulously transform the original data into a Gaussian distribution, remains intact. Consequently, the essential characteristics and distributional properties of the original data are not disrupted. Mathematically, this is formulated as:

\vskip -0.1in
\begin{equation}
	\hat{q}(\mathbf{x}_{1:T} | \mathbf{x}_0) = q(\mathbf{x}_{1:T} | \mathbf{x}_0),
\end{equation}
where $\hat{q}$ denotes the transition probability distribution after introducing the poisoning vector $\mathbf{v}$, and $q$ is the original probability distribution. This invariance is strategically beneficial for an attacker as it ensures that the forward diffusion process can assimilate $\mathbf{v}$ without disturbing the intrinsic structure of the original data, thereby preserving the authenticity of the generated samples.

\textbf{Property 2: Sensitivity of Reverse Sampling}

Conversely, the reverse sampling process is sensitive to the inclusion of $\mathbf{v}$, as it introduces an additional term to the gradient used for data generation. The updated gradient in reverse sampling is given by:
\begin{equation}
	{\nabla _{{\mathbf{x}_t}}}\log p({\mathbf{x}_t}) + {\nabla _{{\mathbf{x}_t}}}\log p(\mathbf{v}\left| {{\mathbf{x}_t}} \right.).
\end{equation}

This adjustment elucidates how the poisoning vector $\mathbf{v}$ engages with intermediate states $\mathbf{x}_t$ during the reverse process, granting fine-grained control over poisoned sample generation. 

\textbf{Property 3: Refined Impact on Poisoned Data}

To quantify and elucidate the impact of $\mathbf{v}$ further, we conduct a comprehensive analysis of its effects on the reverse diffusion trajectory. The modification on the poisoned data $\mathbf{x}_k^p$ is given by:
\vskip -0.15in
\begin{equation}
	\mathbf{x}^{p}_{k}(\mathbf{v}) \approx \mathbf{x}_{k} + \nabla_{\mathbf{v}} \mathbf{x}^{p}_{k} \cdot \mu_v + \frac{1}{2} \sigma_v^2 \mathbf{z}^T \mathbf{H} \mathbf{z} + O(\mathbf{v}^3).
\end{equation}
Here, $\sigma_v$ is responsible for the diversity and fluctuations in the poisoned data, while $\mu_v$ influences the bias introduced to the poisoned dataset.

%\vskip 0.05in
\subsubsection{Jumping Diffusion Strategy} 

PCDM generates effective and stealthy poisoned data without requiring detailed diffusion and estimation. Instead, PCDM adds substantial noise at each time step of forward diffusion, enabling $x_t$ to rapidly approximate a Gaussian distribution and dramatically reducing the diffusion timestep.

``Large Strides'': Compared to the noise variance sequence $\{ \beta _ { t } \in ( 0 , 1 ) \} _ { t = 1 } ^ { T }$ in baseline DDPM, PCDM employs a tailored variance sequence $\{ \hat{\beta} _ { t } \in ( 0 , 1 ) \} _ { t = 1 } ^ { \hat{T} }$, which significantly increases the noise introduced at each time step by utilizing an extended sequence $\{ e _ { t } \} _ { t = 1 } ^ { \hat{T} }$ during the forward diffusion process. At time step $t$, the noise variance with a large stride is $\hat{\beta} _ { t }= e _ { t }{\beta} _ { t },t\in[1,\hat{T}]$. 

This modification accelerates the noise addition while disrupting the equilibrium between the forward and reverse diffusion processes, making it challenging for the reverse process to accurately predict the noise term and leading to insufficient denoising accuracy. Consequently, this approach ensures that the harmful features of the poisoned data are effectively preserved.

``Few Steps'': As a lightweight module designed for generating poisoned data, PCDM significantly reduces the demand for computational resources and time. It efficiently utilizes several time steps to produce a large volume of poisoned data. To ensure the generation of high-quality poisoned data while maintaining a lightweight operation, the total number of time steps $\hat{T}$ in PCDM is set in accordance with the variance sequence $\{ \hat{\beta}_{t} \}$. Specifically, $\hat{T} = T / \bar{e}$, where: 
\vskip -0.05in
\begin{equation}
	\bar{e}=\frac{\varSigma _{t=1}^{\hat{T}}e_t}{\hat{T}}.
\end{equation} 

In practice, $T \geq 1000$ is typically set to ensure that the final noised sample $\mathbf{x}_T$ closely approximates pure noise. Therefore, in PCDM, as the expansion sequence $\{ \hat{\beta} _ { t } \in ( 0 , 1 ) \} _ { t = 1 } ^ { \hat{T} }$ increases, the total number of time steps $\hat{T}$ decreases, i.e., ``larger strides, fewer steps''. This leads to a coarser diffusion process, making the generated poisoned data vaguer and more harmful. Conversely, the opposite holds.

\subsection{Theoretical Analysis}

\subsubsection{The Forward Diffusion Process in PCDM (Property 1)}

The diffusion model is a Markov chain, defined as follows: The state $\mathbf{x}{}_t$ at any given time depends only on the previous time step. Therefore, the conditional probability of the forward diffusion process (noise addition) at any single time step must be independent of $\mathbf{v}$, i.e.,
\vskip -0.1in
\begin{equation}
	\hat q({\mathbf{x}_{1:T}}\left| {\mathbf{x}{}_0} \right.) =q({\mathbf{x}_{1:T}}\left| {\mathbf{x}{}_0} \right.).
\end{equation} 
\vskip -0.05in
Building on this fact, another equation can be derived:
\vskip -0.1in
\begin{equation}
	\begin{split}
		\hat q(\mathbf{x}{}_t\left| {\mathbf{x}{}_{t - 1}} \right.) = \int_\mathbf{v} {\hat q(\mathbf{x}{}_t,\mathbf{v}\left| {\mathbf{x}{}_{t - 1}} \right.)d\mathbf{v}} \\
		= \int_\mathbf{v} {\hat q(\mathbf{x}{}_t\left| {\mathbf{v},\mathbf{x}{}_{t - 1}} \right.)\hat q(\mathbf{v}\left| {\mathbf{x}{}_{t - 1}} \right.)d\mathbf{v}} \\
		= \int_\mathbf{v} {q(\mathbf{x}{}_t\left| {\mathbf{x}{}_{t - 1}} \right.)\hat q(\mathbf{v}\left| {\mathbf{x}{}_{t - 1}} \right.)d\mathbf{v}} \\
		= q(\mathbf{x}{}_t\left| {\mathbf{x}{}_{t - 1}} \right.)\underbrace {\int_\mathbf{v} {\hat q(\mathbf{v}\left| {\mathbf{x}{}_{t - 1}} \right.)d\mathbf{v}} }_{ = 1}\\
		= q(\mathbf{x}{}_t\left| {\mathbf{x}{}_{t - 1}} \right.)
		= \hat q(\mathbf{x}{}_t\left| {\mathbf{x}{}_{t - 1},\mathbf{v}} \right.).
	\end{split}
\end{equation} 
% %\vskip -0.05in
During the forward diffusion process, noise is progressively added in the form of Gaussian perturbations, and the transition probability $q(\mathbf{x}_t | \mathbf{x}_{t-1})$ depends solely on the noise level at time step $t$, which is determined by the noise schedule $\beta_t$. After the introduction of $\mathbf{v}$: (1) $\mathbf{v}$ serves as additional contextual information, but its influence is limited to $\hat{q}(\mathbf{v} | \mathbf{x}_{t-1})$. (2) Due to the normalization property of $\hat{q}(\mathbf{v} | \mathbf{x}_{t-1})$, $\mathbf{v}$ exerts no observable impact on the specific transition process. 

Following the same logic, it can be derived that the joint probability $\hat q({\mathbf{x}_{1:T}}\left| {\mathbf{x}{}_0} \right.)$of the forward diffusion process in the conditional diffusion model is equivalent to that $q({\mathbf{x}_{1:T}}\left| {\mathbf{x}{}_0} \right.)$ in the unconditional (original) diffusion model:
%\vskip -0.1in
\begin{equation}
	\begin{split}
		\hat q({\mathbf{x}_{1:T}}\left| {\mathbf{x}{}_0} \right.) = \int_\mathbf{v} {\hat q({\mathbf{x}_{1:T}},\mathbf{v}\left| {\mathbf{x}{}_0} \right.)d\mathbf{v}} \\
		= \int_\mathbf{v} {\hat q(\mathbf{v}\left| {\mathbf{x}{}_0} \right.)\hat q({\mathbf{x}_{1:T}}\left| {\mathbf{x}{}_0,\mathbf{v}} \right.)d\mathbf{v}} \\
		= \int_\mathbf{v} {\hat q(\mathbf{v}\left| {\mathbf{x}{}_0} \right.)\prod\limits_{t = 1}^T {\hat q({\mathbf{x}_t}\left| {\mathbf{x}{}_{t - 1},\mathbf{v}} \right.)} d\mathbf{v}} \\
		= \int_\mathbf{v} {\hat q(\mathbf{v}\left| {\mathbf{x}{}_0} \right.)\prod\limits_{t = 1}^T {\hat q({\mathbf{x}_t}\left| {\mathbf{x}{}_{t - 1}} \right.)} d\mathbf{v}} \\
		= \prod\limits_{t = 1}^T {\hat q({\mathbf{x}_t}\left| {\mathbf{x}{}_{t - 1}} \right.)} \underbrace {\int_\mathbf{v} {\hat q(\mathbf{v}\left| {\mathbf{x}{}_0} \right.)d\mathbf{v}} }_{ = 1}\\
		= \prod\limits_{t = 1}^T {\hat q({\mathbf{x}_t}\left| {\mathbf{x}{}_{t - 1}} \right.)} 
		= q({\mathbf{x}_{1:T}}\left| {\mathbf{x}{}_0} \right.).
	\end{split}
\end{equation} 

Consequently, $\mathbf{v}$ does not alter the forward diffusion process. As the forward diffusion process uniformly transforms the data into a Gaussian distribution, the addition of global context noise (particularly unbiased noise) does not systematically disrupt the core features of the original data.

\subsubsection{The Backward Diffusion Process in PCDM (Property 2)}

The joint probability of the reverse process in the baseline diffusion model is given by:
%\vskip -0.1in
\begin{equation}
	p(\mathbf{x}{}_{0:T}) = p(\mathbf{x}{}_T)\prod\limits_{t = 1}^T {{p_\theta }({\mathbf{x}_{t - 1}}\left| {{\mathbf{x}_t}} \right.)} .
\end{equation} 

After introducing$\mathbf{v}$, the joint probability of the reverse process in PCDM is given by:
%\vskip -0.05in
\begin{equation}
	p(\mathbf{x}{}_{0:T}\left| \mathbf{v} \right.) = p(\mathbf{x}{}_T)\prod\limits_{t = 1}^T {{p_\theta }({\mathbf{x}_{t - 1}}\left| {{\mathbf{x}_t},\mathbf{v}} \right.)} .
\end{equation} 

In this paper, we analyze the denoising process of the diffusion model from a score-based perspective $\hat{s}_{\theta}(\cdot)$, where the parameterized neural network predicts the logarithmic gradient of $\mathbf{x}_t$, as expressed by the following equation:
\[{\hat {s}_\theta }({\mathbf{x}_t},t) \approx {\nabla _{{\mathbf{x}_t}}}\log p({\mathbf{x}_t}).\]
After introducing $\mathbf{v}$, the original ${\nabla _{{\mathbf{x}_t}}}\log p({\mathbf{x}_t})$ becomes ${\nabla _{{\mathbf{x}_t}}}\log p({\mathbf{x}_t}\left| \mathbf{v} \right.)$:
\begin{equation}
	\begin{split}
		{\nabla _{{\mathbf{x}_t}}}\log p({\mathbf{x}_t}\left| \mathbf{v} \right.) = {\nabla _{{\mathbf{x}_t}}}\left( {\frac{{p({\mathbf{x}_t})p(\mathbf{v}\left| {{\mathbf{x}_t}} \right.)}}{{p(\mathbf{v})}}} \right)\\
		= {\nabla _{{\mathbf{x}_t}}}\log p({\mathbf{x}_t}) + {\nabla _{{\mathbf{x}_t}}}\log p(\mathbf{v}\left| {{\mathbf{x}_t}} \right.) - \underbrace {{\nabla _{{\mathbf{x}_t}}}\log p(\mathbf{v})}_{ = 0}\\
		= {\nabla _{{\mathbf{x}_t}}}\log p({\mathbf{x}_t}) + {\nabla _{{\mathbf{x}_t}}}\log p(\mathbf{v}\left| {{\mathbf{x}_t}} \right.).
	\end{split}
\end{equation} 
${\nabla _{{\mathbf{x}_t}}}\log p({\mathbf{x}_t})$ represents the gradient of the original diffusion model, referred to as the $unconditional$ $score$, and ${\nabla _{{\mathbf{x}_t}}}\log p(\mathbf{v}\left| {{\mathbf{x}_t}} \right.)$ is known as the $adversarial$ $gradient$, where $p(\mathbf{v}\left| {{\mathbf{x}_t}} \right.)$ defines the poisoning-conditional probability distribution, and $\mathbf{v}$ is closely related to the diversity and uncertainty of the poisoned samples.

\subsubsection{The Poisoned Data Generation in PCDM (Property 3)}

Expanding the poisoned output $\mathbf{x}^{p}_{k}$ using a Taylor series around $\mathbf{v} = 0$:
\begin{equation}\label{eq:12}
	\mathbf{x}^{p}_{k}(\mathbf{v}) \approx \mathbf{x}_{k} + \nabla_{\mathbf{v}} \mathbf{x}^{p}_{k} \cdot \mathbf{v} + \frac{1}{2} \mathbf{v}^T  \mathbf{H} \mathbf{v},
\end{equation} 
where $\mathbf{x}_{k}$ is the baseline output (no poisoning is added), $ \nabla_{\mathbf{v}} \mathbf{x}^{p}_{k} $ is the gradient of the output with respect to the noise, indicating the sensitivity of the output to poisoning perturbations, and $ \mathbf{H} $ is the Hessian matrix, representing the second-order effect of poisoning on the output. If $\sigma_v$ is small, the influence of noise primarily manifests in the linear term, $ \nabla_{\mathbf{v}} \mathbf{x}^{p}_{k} \cdot \mathbf{v}$. However, as the poisoning magnitude increases, the quadratic term $ \frac{1}{2} \mathbf{v}^T \mathbf{H} \mathbf{v} $ becomes significant, and the poisoned data is more strongly influenced by the poisoning vector.

Next, we analyze the impact of $\mathbf{v}\sim\mathcal{N}(\mu_v, \Sigma_v)$ on generating poisoned data, where $\mu_v$ is the mean, $\Sigma_v$ is the covariance matrix, and $\sigma_v = \text{diag}(\Sigma_v)$ is the standard deviation. The components of the poisoning vector can be represented as $\mathbf{v} = \sigma_v \cdot \mathbf{z}$, where $\mathbf{z}$ follows a standard Gaussian distribution.

\textbf{Mean term:} The effect of the mean $\mu_v$ is linear and directly reflected in the gradient term:
\[
\nabla_{\mathbf{v}} \mathbf{x}^{p}_{k} \cdot \mu_v.
\]

\textbf{Variance Term:} The variance term is closely related to the standard deviation $\sigma_v$. The magnitude of the poisoning vector is influenced by $\sigma_v$, which in turn affects the second-order term in the generated result. The covariance matrix of $\mathbf{v}$ is $\Sigma_v = \sigma_v^2 \mathbf{I}$, where $\mathbf{I}$ is the identity matrix. For each component of the poisoning vector with standard deviation $\sigma_v$, the effect of the variance is modeled through the Hessian matrix term $\frac{1}{2} \mathbf{v}^T \mathbf{H} \mathbf{v}$, which governs the second-order derivative terms. Therefore, the expression for the effect of the poisoning vector's standard deviation is:
\begin{equation}
	\mathbf{v}^T \mathbf{H} \mathbf{v} = \sigma_v^2 \mathbf{z}^T \mathbf{H} \mathbf{z},
\end{equation} 
where $\mathbf{z} \sim \mathcal{N}(0, \mathbf{I})$ is the standard normal noise vector, and $\sigma_v^2$ is the variance of the poisoning vector (the variance of each component). Thus, the impact of the standard deviation can be reflected through this expression:
\begin{equation}
	\frac{1}{2} \mathbf{v}^T \mathbf{H} \mathbf{v} = \frac{1}{2} \sigma_v^2 \mathbf{z}^T \mathbf{H} \mathbf{z}.
\end{equation} 

After incorporating both the standard deviation $\sigma_v$ and the mean $\mu_v$ into Equation~\ref{eq:12}, the final formula becomes:
%\vskip -0.1in
\begin{equation}
	\mathbf{x}^{p}_{k}(\mathbf{v}) \approx \mathbf{x}_{k} + \nabla_{\mathbf{v}} \mathbf{x}^{p}_{k} \cdot \mu_v + \frac{1}{2} \sigma_v^2 \mathbf{z}^T \mathbf{H} \mathbf{z} + O(\mathbf{v}^3),
\end{equation}
where $\mathbf{x}_{k}$ represents the case without poisoning vector, $\nabla_{\mathbf{v}} \mathbf{x}^{p}_{k} \cdot \mu_v$ and $\frac{1}{2} \sigma_v^2 \mathbf{z}^T \mathbf{H} \mathbf{z}$ respectively represent the influence of the poisoning vector's mean and standard deviation on the generated result, affecting the fluctuations and diversity of the generation, and $O(\mathbf{v}^3)$ represents higher-order terms, which are typically negligible.

In summary, the standard deviation $\sigma_v$ controls the diversity and fluctuations of the poisoned data by influencing the second-order derivative terms (the quadratic terms of the Hessian matrix). On the other hand, the effect of the mean $\mu_v$ is reflected through the first-order derivative terms (the gradient), controlling the bias of the poisoned data.

\subsubsection{Refining the Impact of Hyperparameters $\{ e_{t} \}$ and $\hat{T}$ on PCDM (Property 3)}

During DDPM training, the model aims to minimize the loss by reverse diffusion, gradually recovering a clear image from noise. For simplicity, the loss function is typically computed based on the Euclidean distance between images. However, Euclidean distance may only effectively measure the actual quality of the generated images if the input and output images are very close, which can produce more precise results. Therefore, selecting the most significant possible time step helps make the input and output images as close as possible, thereby mitigating the blurring effect caused by the Euclidean distance.

Therefore, in Equation~\ref{eq:6}, there should be $\bar{\alpha}_T\approx0$. The overall product of \( \alpha_t \) up to step \( T \) is $\bar{\alpha}_T = \prod_{t=1}^T \alpha_t$. Based on the derivations in \cite{ho2020denoising}, the value of $\bar{\alpha}_T$ can be estimated. Substituting the given form of \( \alpha_t \):
\vskip -0.1in
\begin{equation}
	\log \bar{\alpha}_T = \sum_{t=1}^T \log \alpha_t.
\end{equation}
\vskip -0.05in
Approximating \( \log \alpha_t \) for large \( T \):
\begin{equation}
	\begin{split}
		\log \alpha_t = &\log \left( 1 - \frac{0.02 t}{T} \right)^{1/2}\\ 
		< &\log \left( - \frac{0.02 t}{T} \right)^{1/2}.\\
	\end{split}
\end{equation}
%\vskip -0.05in
Thus:
\begin{equation}
	\log \bar{\alpha}_T \approx -0.005 (T+1).
\end{equation}
When \( T = 1000 \), $ \bar{\alpha}_T \approx e^{-5}$, this value is nearly 0, ensuring that the final noised sample $ \mathbf{x}_T $ closely approximates pure Gaussian noise. Choosing a sufficiently large $T$ ensures that adequate noise is introduced, which not only facilitates the diffusion process but also promotes greater diversity in the generated poisoned data while maintaining its core features.

However, PCDM aims to generate specific poisoned data rather than refine the diffusion process to produce precise results. Therefore, PCDM focuses on the noise variance sequence $\{ \hat{\beta} _ { t }\}_{t=1}^{\hat{T}}$ to satisfy the condition. It is known that:
\vskip -0.1in
\begin{equation}
	\log \bar{\alpha}_T = \sum_{t=1}^{\hat{T}} \log(1 - \beta_t).
\end{equation}

For example, if $\hat{\beta_t}\in\{ \hat{\beta} _ { t }\}\approx 0.1$, we can estimate $\log(1 - 0.1) \approx -0.105$. Thus, $\log \bar{\alpha}_{\hat{T}} \approx -0.105\hat{T}$. In order for $\alpha_T \approx 0$, $\log \bar{\alpha}_{\hat{T}} \approx -5$ is required, which gives $\hat{T} \approx \frac{5}{0.105} \approx 48$. This means that when $\beta_t$ is large, $\hat{T}$ can be chosen to be a smaller value. This choice ensures that after $\hat{T}$ steps, $\alpha_{\hat{T}}$ becomes sufficiently small, thereby allowing the noise introduction process to fully unfold. In PCDM,
\begin{equation}
	\hat{\beta} _ { t }=e_t\cdot{\beta} _ { t },\hat{\beta} _ { t }\in\{\hat{\beta} _ { t }\},e_t\in\{e_t\},{\beta} _ { t }\in\{{\beta} _ { t }\}_{t=1}^{\hat{T}}.
\end{equation}

PCDM needs to ensure that the accumulated noise increments under its time steps $\hat{T}$ are consistent with baseline time steps $T$. This condition can be expressed by accumulating the noise increments. Baseline accumulated noise increment:
\vskip -0.1in
\begin{equation}
	\bar{\beta}_T = \sum_{t=1}^{T} \beta_t.
\end{equation}
\vskip -0.05in
PCDM accumulated noise increment:
\begin{equation}
	\bar{\hat{\beta}}_{\hat{T}} = \sum_{t=1}^{\hat{T}} \hat{\beta}_t = \sum_{t=1}^{\hat{T}} e_t \cdot \beta_t = \bar{e} \sum_{t=1}^{\hat{T}} \beta_t.
\end{equation}
\vskip -0.05in
To ensure a smooth and stable noise addition and removal process across different time steps, PCDM must maintain consistency in the cumulative noise schedule with the baseline DDPM. This consistency ensures theoretical soundness by preserving both the smoothness of per-step noise addition and the learnability of the denoising process, ultimately allowing the core features of the data to remain intact. To ensure theoretical equivalence with the baseline, we impose the condition: $\bar{\beta}_T = \bar{\hat{\beta}}_{\hat{T}}$. That is,
\vskip -0.15in
\begin{equation}
	\sum_{t=1}^{T} \beta_t = \bar{e} \sum_{t=1}^{\hat{T}} \beta_t.
\end{equation}
\vskip -0.05in
By aligning $\bar{\beta}_T$ with $\bar{\hat{\beta}}_{\hat{T}}$, we ensure that the noise addition at each step remains smooth and comparable. Additionally, the smoothness in $\beta_t$ ensures that the perturbation introduced to $\mathbf{x}_t$ does not disrupt the core features of the data, making the backward diffusion process (denoising) both practical and learnable for PCDM.

Furthermore, preserving the cumulative noise schedule has two key benefits: (1) Smooth Noise Transition: By maintaining a consistent noise trajectory, each intermediate step $t$ transitions seamlessly between states. This prevents abrupt variations in noise levels that could impair feature preservation and destabilize the reverse process. (2) Denoising Learnability: A smooth and consistent noise schedule ensures that the model can effectively approximate the denoising distribution $p_\theta(\mathbf{x}_{t-1}|\mathbf{x}_t)$ at each step, leading to better reconstructions of the poisoned data while retaining its core characteristics.

For the case where $\beta_t$ increases linearly, $\hat{T}$ and $T$ should satisfy a proportional relationship:
\begin{equation}
	\hat{T} = \frac{T}{\bar{e}}.
\end{equation}
Substituting $T=1000$ yields $\hat{T} = \frac{1000}{\bar{e}}$, indicating that as the noise schedule parameters increase, the number of time steps needs to decrease to maintain the same total noise increment.

\subsection{Diffusion-based Poisoned Data Generation}

Given a malicious client $c_j$ with a dataset $\mathcal{D}_j$, the attacker uses a PCDM $G$ to generate poisoned dataset $\mathcal{D}_j^p$, with which the malicious client trains the local poisoned model.  
\vskip -0.1in
\begin{equation}
	\mathcal{D}^{p}_{j}=G\left( \mathbf{\epsilon }_{\theta}, \mathcal{D}_j,\mathbf{v}\right) ,
\end{equation} 
\vskip -0.05in
\noindent where $\epsilon _{\theta}$ is a function approximator intended to predict diffusion noise $\epsilon$, and $\mathbf{v}$ is a global poisoning vector. $G$ generates poisoned data $\mathbf{x}_j^p$ through two stages: training and generation.

\textbf{Training Stage:} 
The objective of the training stage is to optimize the model $G$ by learning the parameters of the noise prediction function $\epsilon_{\theta}$. During training, the gradient of the objective function is computed as follows:

\vskip -0.15in
\begin{equation}
	\begin{split}
		\nabla G&\left( \epsilon _{\theta},\left( \mathbf{x}_k,\mathbf{y}_k,\mathbf{v}\right) _j ,t,e\right) \\
		&,k\in [1,K_j],t\in [0,\hat{T}],e\in [1,E],
	\end{split}
\end{equation} where $\left( \mathbf{x}_j,\mathbf{y}_j \right)\in \mathcal{D}_j$, and $E$ is the training epoch. For each $e=0,1,2,...,E$, $t=0,1,...,\hat{T}-1,\hat{T}$, a gradient descent step is taken according to the PCDM:
\vskip -0.15in

\begin{equation}
	\nabla _{\theta}\lVert \mathbf{\epsilon }-\mathbf{\epsilon }_{\theta}\left( \sqrt{\bar{\alpha}_t}(\mathbf{x}_{j}+r(\mathbf{v}))+\sqrt{1-\bar{\alpha}_t}\mathbf{\epsilon ,}t \right) \rVert ^2.
\end{equation}
%\vskip -0.05in

The loss function learns to denoise the corrupted input by measuring the deviation between the actual noise $\mathbf{\epsilon}$ and the noise predicted by $\epsilon_{\theta}$. After the training stage, $G$ captures the noise patterns introduced during the forward diffusion process, including the influence of the injected global context $\mathbf{v}$. This enables the model to generate poisoned samples that preserve the original data's structure while embedding adversarial perturbations into the generated results.

\textbf{Generation Stage:}
After the training stage, $G$ generates the corresponding poisoned data by starting from random noise $\mathbf{x}_{\hat{T}} \sim \mathcal{N}(0, \mathbf{I})$ and conditioning on the sample label $(\mathbf{y}^{p}_{k})_j$. The model iteratively denoises the input over $\hat{T}$ time steps to generate the poisoned data. A noise term $\mathbf{z}$ is added to achieve a random sample, and $\sigma _t$ is the variance of $\mathbf{z}$.
%\vskip -0.2in
\begin{equation}
	\begin{split}
		\mathbf{x}_{t-1}=&\frac{1}{\sqrt{\alpha _t}}\left( \mathbf{x}_t+r(\mathbf{v})-\frac{1-\alpha _t}{\sqrt{1-\bar{\alpha}_t}}\mathbf{\epsilon }_{\theta}\left( \mathbf{x}_t+r(\mathbf{v}),t \right) \right)\\
		&+\sigma _t\mathbf{z}, \mathbf{z} \sim \mathcal{N}(0,\mathbf{I}).
	\end{split}
\end{equation}
\vskip -0.05in

At each time step $t=\hat{T},\hat{T}-1,...,1,0$, $G$ samples $\mathbf{x}_t$. After sampling $\hat{T}$ time steps, $\mathbf{x}_{0}$ is the poisoned data $\mathbf{x}^{p}_{j}$ generated by PCDM. $G$ constructs the poisoned dataset $\mathcal{D}_j^p=\{(\mathbf{x}^{p}_k,\mathbf{y}_k^p)_{j}\}_{k =1}^{K_p}$ by
\begin{equation}
	\mathcal{D}_{j}^{p}=G\left( \epsilon _{\theta},\left\{ (\mathbf{y}^{p}_{k})_j\right\},\mathbf{v},\mathbf{z,}t \right) ,k\in [1,K_p],t\in [0,\hat{T}],
\end{equation}
where $K_p$ represents the number of poisoned data, and label $( \mathbf{y}_k^p ) _j$ is copied from $( \mathbf{y}_k ) _j$. 

\subsection{Poisoned Federated Training} 

Every malicious client $c_j$ generates a poisoned dataset $\mathcal{D}^{p}_{j}$ using $G$ and local dataset $\mathcal{D}_{j}$ before federated training, then $\mathcal{D}^{p}_{j}$ is used for poisoned training instead of $\mathcal{D}_{j}$. In the initial round, the malicious client trains a poisoned local model $\omega _{j}^{*\left( 1 \right)}$ with poisoned dataset $\mathcal{D}^{p}_{j}$.
% %\vskip -0.2in
\vskip -0.05in
\begin{equation}
	\omega _{j}^{*\left( 1 \right)}=\nabla F\left( \mathcal{D}_{j}^{p} \right) .
\end{equation}
\vskip -0.05in
After that, $c_j$ uploads the poisoned local model $\omega _{j}^{*\left( 1 \right)}$ to the server, and federated aggregation will result in a poisoned global model $\hat{\omega} _{g}^{\left( 1 \right)}$ if server selects the poisoned model $\omega _{j}^{*\left( 1 \right)}$.

In the following rounds, the malicious client receives a poisoned global model $\hat{\omega}_{g}^{\left( r \right)}$, trains a new poisoned local model $\omega _{j}^{*\left( r+1 \right)}$ based on $\hat{\omega}_{g}^{\left( r \right)}$ using poisoned dataset $\mathcal{D}_j^p$, and uploads it to the central server.
\begin{equation}
	\omega _{j}^{*\left( r+1 \right)}=\hat{\omega}_{g}^{\left( r \right)}-\eta \cdot \nabla F\left( \hat{\omega} _{g}^{\left( r \right)},\mathcal{D}_j^p \right) .
\end{equation}
%\vskip -0.1in
  The malicious client performs receiving, training, and uploading in such a loop.

\subsection{Implementation Guidelines} \label{trade}

If a poisoned local model is not detected as abnormal by the server, it can undermine the global model during federated aggregation. Since detectable poisoning is readily neutralized, PCDM is designed to impose maximal aggregation damage while maintaining benign update profiles%Therefore, the use of PCDM should consider both attack effectiveness and stealthiness.

\subsubsection{Attack Effectivieness}\label{guide}

The test accuracy of the poisoned global model reflects the effectiveness of data poisoning attacks. PCDM aims for the optimal poisoned dataset $\mathcal{D}_{j}^{p}$ to indirectly undermine the global model, thereby  maximizing the reduction in test accuracy of global model. 

\vskip -0.15in
\begin{equation}
	\underset{\mathcal{D}_{j}^{p}}{\max} \quad a_r-\hat{a}_r,
\end{equation}
\vskip -0.03in
\noindent where, $a_r=F\left( \omega _{g}^{\left( r \right)},\mathcal{D}_{test} \right)$, $\hat{a}_r=F\left( \hat{\omega} _{g}^{\left( r \right)},\mathcal{D}_{test} \right)$, $\hat{\omega} _{g}^{\left( r \right)}$ and $\omega _{g}^{\left( r \right)}$ denote the poisoned and unpoisoned global models.

\subsubsection{Attack Stealthiness}\label{stea}

The server typically defends against data poisoning attacks by calculating the difference between each local model and other models:
% %\vskip -0.2in
\begin{equation}
	\sum_{i=1}^B{\frac{\lVert \phi \left( \omega _{j}^{*\left( r \right)} \right) ,\phi \left( \omega _{i}^{\left( r \right)} \right) \rVert}{B}},
\end{equation}
\vskip -0.05in
\noindent then discard local models with large differences. The low-dimensional kernel function $\phi(\cdot)$ (e.g. PCA, KPCA, SVD) is used to extract critical features of local models, as computing differences between high-dimensional models is challenging and costly due to the massive number of parameters.

Reviewing local models at each round is time-consuming and computationally expensive. Furthermore, highly dense defenses can trigger many false alarms. Therefore, previous studies have employed an interval defense method by storing the local models received during an interval of $R'$ time steps to construct a model set for each client, followed by calculating the differences between these local model sets (e.g., using Euclidean distance or cosine similarity).
%\vskip -0.2in
\begin{equation}
	s_j^{(R')}=\sum_{i=1}^N{\frac{\lVert \left\{ \phi \left( \omega _{j}^{*\left( r \right)} \right) \right\} _{r=1}^{R'},\left\{ \phi \left( \omega _{i}^{\left( r \right)} \right) \right\} _{r	=1}^{R'} \rVert}{N}}.
\end{equation}
\vskip -0.05in
The threshold $S^{(R')}$ or model differences is then set by statistical methods, such as calculating the mean or median of the differences as the threshold. Clients whose $s_j^{(R')}>S^{(R')}$ are labeled as malicious ones. A smaller $s_j^{(R')}$ indicates greater stealthiness and it requires that $s_j^{(R')}<S^{(R')}$.

\subsubsection{Trade-off}

To summarize, PCDM attack can be formulated as the following optimization problem:

\vskip -0.15in
\begin{equation}
	\begin{split}
		\underset{\left\{  \mathcal{D}_j^p\right\}}{\max}& \quad a_r-\hat{a}_r, r\in [1,R] \\
		s.t. &\quad s_j^{(R')}<S^{(R')},j=1,2,...,M.
	\end{split}
\end{equation}
\vskip -0.02in
%The effectiveness and stealthiness of data poisoning attacks are mutually exclusive, meaning that there is typically a trade-off between them. 
%Typically, attack impact and imperceptibility share an inherent trade-off, where increasing one often compromises the other. PCDM does not interfere with the training process of the local model; it controls the poisoned data, thereby achieving a trade-off between attack effectiveness and stealthiness. 

Typically, attack impact and imperceptibility share an inherent trade-off, where increasing one often compromises the other. PCDM addresses this tension by decoupling data manipulation from the local training process, allowing effective model degradation under strict detectability constraints. The former is proportional to the extent of poisoning characteristics, while the latter is proportional to how closely the poisoned data resembles real data.characteristics.

PCDM provides several hyperparameters that offer fine-grained control over the generation of poisoned data, allowing for a trade-off between the effectiveness of the data poisoning attack and its stealthiness:

\textbf{Training epoch $E$:} This primarily affects the extent to which poisoned data aligns with real data. As $E$ increases, the fundamental features in the poisoned data become more distinct, enhancing the stealthiness of the attack. Conversely, a smaller $E$ produces vaguer features, which can reduce the attack's stealthiness and requires fewer computational resources and less time. Therefore, from a cost-effectiveness standpoint, a smaller $E$ is generally preferred when similar attack outcomes are achievable.

\textbf{$\sigma_v$ and $\mu_v$ of poisoning vector:} These two parameters represent the standard deviation and mean of the poisoning vector. Specifically, $\sigma_v$ controls the magnitude of the poisoned features in the data, while $\mu_v$ determines the degree of offset between the poisoned and real data. Both parameters primarily affect the effectiveness of attack: as $\sigma_v$ and $\mu_v$ increase, the attack becomes more effective.

\textbf{Expanded sequence $\{ e _ { t } \}$ and Time step ${ \hat{T} }$:} $\{e_{t}\}$ and ${\hat{T}}$ must satisfy Equation (15) to ensure the proper functioning of PCDM. $\{e_{t}\}$ directly controls the diffusion stride, indirectly determining ${\hat{T}}$. They jointly govern the granularity of the diffusion process: as $\{e_{t}\}$ increases, the corresponding ${\hat{T}}$ decreases, making the process coarser. This results in a more effective but less stealthy attack. Similar to $E$, smaller ${\hat{T}}$ values should be prioritized for the same attack outcomes.

There are various choices for $\{e_{t}\}$. For example, $\{e_{t}\}$ can follow linear sequences (arithmetic progression) or non-linear sequences (exponential progression). Regarding arrangement, options include increasing, decreasing, constant sequences, etc. This means that even for the same ${\hat{T}}$, PCDM offers a variety of diffusion strategies, accompanied by a diverse range of possible poisoned data forms. Hence, $\{e_{t}\}$ can be flexibly configured to suit the specific task.

\section{Experiments}
% \begin{table}[h]
%     \centering
%     \caption{Summary of Dataset Statistics and Data Partitioning used in Experiments.}
%     \label{tab:datasets}
%     \resizebox{\columnwidth}{!}{%
%     \begin{tabular}{l|c|c|c|c}
%         \toprule
%         \textbf{Dataset} &  \textbf{Input Shape} & \textbf{Classes} & \textbf{Training Set} & \textbf{Testing Set} \\ 
%         \midrule
%         MNIST &  $1 \times 28 \times 28$ & 10 & 60,000 & 10,000 \\
%         Fashion-MNIST  & $1 \times 28 \times 28$ & 10 & 60,000 & 10,000 \\
%         CIFAR-10  & $3 \times 32 \times 32$ & 10 & 50,000 & 10,000 \\
%         CIFAR-100 & $3 \times 32 \times 32$ & 100 & 50,000 & 10,000 \\
%         \bottomrule
%     \end{tabular}%
%     }
% \end{table}

\subsection{Experimental Setup}

\textbf{System Overview.} To evaluate the proposed PCDM in realistic wireless edge intelligence scenarios, our experimental design follows the configuration of a standard FL system. We simulate a wireless network comprising 100 distributed clients (e.g., IoT nodes or smart cameras). The global training process spans 200 communication rounds. In each round, a fraction of clients ($0.2$, i.e., 20 clients) are randomly selected to execute local training. Specifically, each selected client performs 5 local epochs with a batch size of 128. The global objective $F$ is to collaboratively train Deep Neural Networks (e.g., CNN and ResNet) for image classification, representing a fundamental application in next-generation wireless networks.

\textbf{Datasets and Data Partitioning.} We utilize five benchmark datasets to represent different task complexities: MNIST \cite{deng2012mnist} and Fashion-MNIST \cite{xiao2017fashion} represent lightweight sensing tasks, while CIFAR-10 and CIFAR-100 \cite{krizhevsky2009learning} represent complex environmental monitoring tasks. The specific statistics and strict data partitioning details for each dataset are summarized in Table~\ref{tab:datasets}. To comprehensively evaluate the robustness of our method against statistical heterogeneity, we simulate two distinct data distribution settings. In the IID setting, the training data is uniformly shuffled and randomly assigned to clients to simulate a balanced environment. In the Non-IID setting, we adopt the data partitioning protocol described in \cite{marfoq2021federated} to simulate realistic statistical heterogeneity where local data is modeled as a mixture of underlying distribufivetions. Specifically, we partition the training data among clients using a symmetric Dirichlet distribution with a concentration parameter $\alpha=0.5$. This setting generates highly skewed label distributions across clients, effectively capturing the inherent heterogeneity of local data distributions. In all scenarios, the testing set remains on the server for global performance evaluation. Crucially, to ensure a fair comparison, the exact same data partitioning indices, random seeds, and client data assignments were applied across all evaluated techniques.

\begin{table}[h]
\vskip -0.15in
    \centering
    \caption{Summary of Dataset Statistics and Data Partitioning.}
    \vskip -0.05in
    \label{tab:datasets}
    \resizebox{\columnwidth}{!}{%
    \begin{tabular}{l|c|c|c|c}
        \toprule
        \textbf{Dataset} &  \textbf{Input Shape} & \textbf{Classes} & \textbf{Training Set} & \textbf{Testing Set} \\ 
        \midrule
        MNIST &  $1 \times 28 \times 28$ & 10 & 60,000 & 10,000 \\
        Fashion-MNIST  & $1 \times 28 \times 28$ & 10 & 60,000 & 10,000 \\
        CIFAR-10  & $3 \times 32 \times 32$ & 10 & 50,000 & 10,000 \\
        CIFAR-100 & $3 \times 32 \times 32$ & 100 & 50,000 & 10,000 \\
        % VRAI & $3 \times 64 \times 64$ & 7 & 52,890 & 13,223 \\
        \bottomrule
    \end{tabular}%
    }
    \vskip -0.05in
\end{table}

\textbf{Model Architectures.} We deploy model architectures tailored to the complexity of the learning tasks. For the lightweight MNIST and Fashion-MNIST datasets, we utilize a compact CNN consisting of two convolutional layers (16 and 32 channels) followed by a fully connected layer. For the more complex CIFAR-10 dataset, we employ a deeper network comprising six convolutional layers structured into three blocks (with 32, 64, and 128 channels, respectively) and two fully connected layers. The model for CIFAR-100 retains the same depth as the CIFAR-10 architecture but doubles the channel width to accommodate the expanded label space. All convolutional layers are equipped with Batch Normalization, ReLU activation, and Max Pooling to ensure training stability and performance.

\textbf{Baselines.} We replicate four kinds of mainstream data poisoning methods to compare with PCDM.
\begin{itemize} 
	\item Label flipping attack (LF)  \cite{tolpegin2020data}: This method changes the labels of specific classes in the training data to perform untargeted poisoning in FL.
	\item PoisonGAN (LF+PoisonGAN) \cite{zhang2020poisongan}: This method applies a data augmentation method using off-the-shelf GANs to enhance attack effectiveness, such as generating pseudo-samples to increase dataset size before flipping labels.
	\item Noise superimposition attack (NS) \cite{gragnaniello2018analysis,yang2023clean}: This method involves adding various types of noise, such as Gaussian (LNS, HNS) or SAP (SAP-NS) noise, directly to real data to create poisoned samples.
	\item VagueGAN \cite{10287523}: This method utilizes specialized variants of GANs to generate vague yet effective poisoned data, thus achieving a balance between attack impact and stealthiness in FL.
\end{itemize}

\textbf{Defense Evaluation.} We utilize seven representative defense methods to evaluate the actual impact of data poisoning attacks in this paper.
\begin{itemize} 
	\item PCA \cite{tolpegin2020data}: This method constructs and standardizes a list of local models for dimensional reduction using PCA and identifies significant outliers.
    \item UMAP \cite{upreti2022defending}: This method employs UMAP for non-linear dimensionality reduction of model gradients, utilizing cosine distance metrics with a neighborhood size of 100 and minimum distance of 0.6. It identifies poisoned models based on Euclidean distance from the centroid. 
    \item CONTRA \cite{awan2021contra}: This method detects poisoned models by calculating the cosine similarity matrices, where models with significantly deviated gradient directions are marked as poisoned models. 
    % \item DnC \cite{shejwalkar2021manipulating}: This method identifies poisoned models by analyzing the principal component direction of gradient updates. It first performs dimensionality reduction through sampling, then calculates the principal components and projects all gradients, and finally eliminating the models with the largest projection values as poisoned models.
    \item  DnC \cite{shejwalkar2021manipulating}:  This method identifies poisoned models by analyzing the principal direction of gradient updates. It samples to reduce dimensionality, computes principal components, projects gradients, and eliminates the models with the largest projections as poisoned ones.
    \item K-Means \cite{li2022robust, onsu2023cope}: This method performs K-Means clustering on the model gradients after dimensionality reduction through PCA to identify abnormal clusters.
    \item FedDMC \cite{mu2024feddmc}: This method employs binary tree-based noise clustering (BTBCN) to detect poisoned models after performing PCA dimensionality reduction.
    \item LoMar \cite{li2021lomar}:This method quantifies the poisoned models by computing the relative distribution difference with neighboring updates using a non-parametric kernel density estimation approach.
    \item MCD \cite{sun2024gan}: This method conducts periodic rigorous reviews of the distribution behavior of local model parameters, which can identify the vast majority of existing data poisoning attacks, including recent GAN-based ones.
\end{itemize}

\textbf{Robust Aggregation Evaluation.} We further compare with three representative Byzantine-robust aggregation rules:
\begin{itemize} 
	\item Multi-Krum \cite{blanchard2017machine}: This distance-based method selects a subset of local models with the smallest sum of squared Euclidean distances to their neighbors, effectively filtering outliers before aggregation.
	\item SignGuard \cite{xu2022byzantine}: This method integrates direction-based clustering using gradient signal statistics with magnitude-based filtering to identify and exclude malicious models exhibiting abnormal statistical behaviors.
	\item LASA \cite{xu2025achieving}: This method employs layer-adaptive sparsified aggregation. It combines pre-aggregation top-$k$ sparsification to reduce the attack surface with a layer-wise filter that selects benign layers based on magnitude and Positive Direction Purity metrics.
\end{itemize}

\subsection{Experimental Result}

\begin{table*}[t]
\caption{Test accuracy in presence of data poisoning attacks  (\%)}
%%\vskip -0.1in
\centering
\setlength{\tabcolsep}{0.04in}  % 减小列间距
\label{tab:my-table}
\begin{tabular}{llccccccccccccccccccccccc}
\hline
\multirow{3}{*}{\begin{tabular}[c]{@{}l@{}}Data \\ Distribution\end{tabular}} & \multirow{3}{*}{Attack} & \multicolumn{23}{c}{Dataset and Malicious Clients Percentage $\alpha$}                                                                                                                                                                                                                                                                                                               \\ \cline{3-25} 
                                                                              &                         & \multicolumn{5}{c}{MNIST}                                            &  & \multicolumn{5}{c}{Fashion-MNIST}                                    &  & \multicolumn{5}{c}{CIFAR-10}                                         &  & \multicolumn{5}{c}{CIFAR-100}                                                                                                                          \\ \cline{3-7} \cline{9-13} \cline{15-19} \cline{21-25} 
                                                                              &                         & 0\%  & 5\%           & 10\%          & 20\%          & 30\%          &  & 0\%  & 5\%           & 10\%          & 20\%          & 30\%          &  & 0\%  & 5\%           & 10\%          & 20\%          & 30\%          &  & 0\%                      & 5\%                      & 10\%                     & 20\%                              & 30\%                              \\ \hline
\multirow{7}{*}{IID}                                                          & Label Flipping          & 98.7 & 98.7          & 98.5          & 98.3          & 98.1          &  & 88.0 & 87.7          & 87.2          & 86.0          & 85.5          &  & 76.3 & 75.7          & 75.0          & 73.9          & 72.9          &  & 55.7                     & 53.0                     & 52.5                     & 49.3                              & 46.7                              \\
                                                                              & PoisonGAN               & 98.7 & 98.6          & 98.4          & 98.2          & 98.0          &  & 88.0 & 87.6          & 87.2          & 85.8          & 85.1          &  & 76.3 & 75.6          & 74.8          & 73.8          & 72.7          &  & 55.7                     & 52.8                     & 52.1                     & 48.4                              & 46.9                              \\
                                                                              & Light Noise             & 98.7 & 98.7          & 98.6          & 98.4          & 98.3          &  & 88.0 & 87.8          & 87.6          & 86.9          & 86.6          &  & 76.3 & 76.0          & 75.7          & 74.9          & 74.1          &  & 55.7                     & 54.0                     & 53.7                     & 51.1                              & 50.0                              \\
                                                                              & Heavy Noise             & 98.7 & 98.6          & 98.5          & 98.4          & 98.1          &  & 88.0 & 87.7          & 87.2          & 86.2          & 85.6          &  & 76.3 & 75.9          & 75.1          & 74.2          & 73.1          &  & 55.7                     & 53.5                     & 52.5                     & 50.2                              & 48.6                              \\
                                                                              & SAP Noise               & 98.7 & 98.7          & 98.6          & 98.6          & 98.4          &  & 88.0 & 87.8          & 87.5          & 87.0          & 86.4          &  & 76.3 & 76.1          & 75.7          & 74.7          & 74.0          &  & 55.7                     & 54.1                     & 53.8                     & 51.2                              & 49.6                              \\
                                                                              & VagueGAN                & 98.7 & 98.6          & 98.4          & 97.8          & 97.5          &  & 88.0 & 87.3          & 86.8          & 85.9          & 84.8          &  & 76.3 & 74.9          & 74.0          & 73.2          & 71.9          &  & 55.7                     & 51.5                     & 48.3                     & \textbf{42.4}                     & 39.4                              \\
                                                                              & PCDM                    & 98.7 & \textbf{98.5} & \textbf{98.4} & \textbf{97.6} & \textbf{97.1} &  & 88.0 & \textbf{87.1} & \textbf{86.5} & \textbf{85.7} & \textbf{84.2} &  & 76.3 & \textbf{74.1} & \textbf{73.8} & \textbf{72.3} & \textbf{71.7} &  & 55.7                     & \textbf{52.1}            & \textbf{47.6}            & 44.7                              & \textbf{37.3}                     \\ \hline
\multirow{7}{*}{non-IID}                                                      & Label Flipping          & 98.4 & 98.3          & 98.1          & 97.9          & 97.6          &  & 87.3 & 86.7          & 86.2          & 85.3          & 84.4          &  & 73.6 & 72.3          & 71.2          & 70.2          & 68.8          &  & \multicolumn{1}{l}{53.4} & \multicolumn{1}{l}{52.1} & \multicolumn{1}{l}{49.3} & \multicolumn{1}{l}{46.8}          & \multicolumn{1}{l}{41.4}          \\
                                                                              & PoisonGAN               & 98.4 & 98.3          & 97.9          & 97.9          & 97.4          &  & 87.3 & \textbf{86.6} & 86.0          & 85.0          & 84.1          &  & 73.6 & 72.2          & 71.1          & 70.0          & 68.7          &  & \multicolumn{1}{l}{53.4} & \multicolumn{1}{l}{51.7} & \multicolumn{1}{l}{49.4} & \multicolumn{1}{l}{46.7}          & \multicolumn{1}{l}{40.0}          \\
                                                                              & Light Noise             & 98.4 & 98.4          & 98.2          & 98.0          & 97.7          &  & 87.3 & 87.1          & 86.8          & 86.3          & 85.7          &  & 73.6 & 72.9          & 72.0          & 71.3          & 70.5          &  & \multicolumn{1}{l}{53.4} & \multicolumn{1}{l}{52.9} & \multicolumn{1}{l}{51.0} & \multicolumn{1}{l}{48.5}          & \multicolumn{1}{l}{46.0}          \\
                                                                              & Heavy Noise             & 98.4 & 98.4          & 98.0          & 97.9          & 97.6          &  & 87.3 & 86.9          & 86.6          & 85.8          & 85.3          &  & 73.6 & 72.8          & 71.9          & 70.7          & 69.9          &  & \multicolumn{1}{l}{53.4} & \multicolumn{1}{l}{51.3} & \multicolumn{1}{l}{48.6} & \multicolumn{1}{l}{45.1}          & \multicolumn{1}{l}{43.0}          \\
                                                                              & SAP Noise               & 98.4 & 98.4          & 98.2          & 98.0          & 97.8          &  & 87.3 & 87.1          & 86.8          & 86.3          & 85.9          &  & 73.6 & 73.0          & 72.0          & 71.2          & 70.4          &  & \multicolumn{1}{l}{53.4} & \multicolumn{1}{l}{51.7} & \multicolumn{1}{l}{50.0} & \multicolumn{1}{l}{48.4}          & \multicolumn{1}{l}{44.7}          \\
                                                                              & VagueGAN                & 98.4 & \textbf{98.1} & 97.9          & 97.8          & 97.4          &  & 87.3 & 86.6          & 85.8          & 84.9          & 84.0          &  & 73.6 & 72.0          & 70.6          & 69.6          & 67.8          &  & \multicolumn{1}{l}{53.4} & \multicolumn{1}{l}{51.0} & \multicolumn{1}{l}{49.0} & \multicolumn{1}{l}{\textbf{41.0}} & \multicolumn{1}{l}{\textbf{35.1}} \\
                                                                              & PCDM                    & 98.4 & 98.2          & \textbf{97.8} & \textbf{97.2} & \textbf{96.9} &  & 87.3 & 86.6          & \textbf{85.0} & \textbf{84.5} & \textbf{83.4} &  & 73.6 & \textbf{71.6} & \textbf{70.0} & \textbf{68.6} & \textbf{66.3} &  & \multicolumn{1}{l}{53.4} & \textbf{49.7}            & \textbf{47.0}            & 41.3                              & 37.2                              \\ \hline
\end{tabular}
\vskip -0.2in
\end{table*}

\textbf{Effectiveness of PCDM Attack.} 
We evaluate the effectiveness of the PCDM and the baseline attacks by measuring the extent of the drop in the global model accuracy caused by each attack in identical FL settings (see Subsection~\ref{guide}).
As illustrated in Table~\ref{tab:my-table}, the impact of each attack method on the global model accuracy is evaluated as the proportion of malicious clients varies from 5\% to 30\%. The results indicate that PCDM consistently demonstrates superior attack effectiveness compared to baseline methods across different tasks, particularly on datasets with high task complexity, like CIFAR-100. Furthermore, all attack methods exhibit greater effectiveness on complex tasks than on simpler ones. Additionally, we investigate the attack performance under non-independent and identically distributed (non-IID) conditions among clients, where PCDM maintains the highest effectiveness.

\textbf{Stealthiness of PCDM Attack.}
To effectively evaluate the stealthiness of PCDM and baseline attacks against defense mechanisms, we employ a visualization strategy to analyze the geometric characteristics of model updates. Since neural network weight updates reside in a high-dimensional space that is fundamentally opaque to direct observation, we utilize Principal Component Analysis (PCA) to project high-dimensional vectors onto a two-dimensional plane. This projection preserves the relative distances and distributional structures of updates, providing an intuitive proxy for how defenses (which typically rely on statistical distance or similarity metrics) perceive benign versus malicious behaviors \cite{tolpegin2020data}. 

\begin{figure}[h]
% \vskip -0.1in
	\begin{center}
	\centerline{\includegraphics[width=\linewidth]{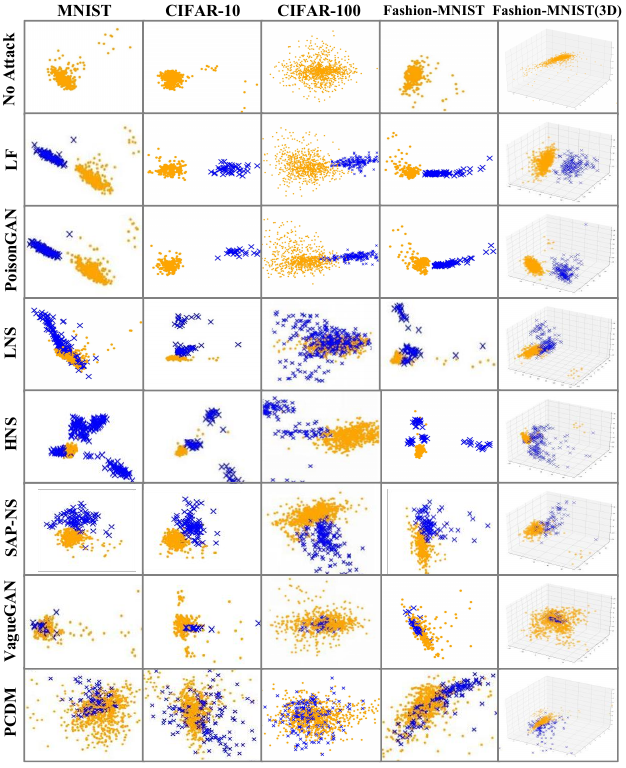}}
    \vskip -0.1in
		\caption{The poisoned models after different attacks.}
		\label{fig:defense}
	\end{center}
    \vskip -0.45in
\end{figure}

As shown in Fig.~\ref{fig:defense}, we present the distribution patterns across four datasets, where yellow `O's denote benign models and blue `X's signify malicious ones. For crude attacks like Label Flipping, the primary distribution of malicious models significantly deviates from that of benign models in terms of Euclidean distance. This distinct geometric separation makes them easily identifiable and filterable by classic outlier detection mechanisms.

However, the visualization of VagueGAN reveals a more complex phenomenon regarding the trade-off between stealthiness and diversity. As observed in Fig.~\ref{fig:defense}, VagueGAN successfully guarantees that malicious updates are embedded within the benign cluster, making them difficult to detect via simple distance measurements. Nevertheless, a critical anomaly emerges: the malicious updates distinctively encompass a much smaller footprint than benign ones. This lack of variance stems from the inherent conceptual limitation of GANs known as ``mode collapse'' \cite{thanh2020catastrophic}. Since VagueGAN is trained via an adversarial min-max game, the generator often fails to capture the full distribution of the target class, resulting in poisoned samples with limited diversity. Consequently, this high consistency in generated data translates into excessively high consistency in model updates, distinctively marking them as anomalous. In contrast, PCDM exhibits superior stealthiness; its malicious updates are indistinguishable from benign ones in both location and distribution spread, confirming that the diffusion-based approach effectively overcomes the mode collapse issue.

To provide a unified and interpretable evaluation of defense performance against poisoning attacks, we define a \textbf{Composite Score} that balances two critical aspects: the ability to correctly identify poisoned models and the capacity to avoid misclassifying benign ones. Specifically, the composite score is computed as a weighted sum of the \textit{Recall} and the \textit{false positive rate (FPR)}, formulated as:
\vskip -0.25in
\begin{equation}
\textbf{Composite Score} = 100 \times (\alpha \cdot \text{Recall} + \beta \cdot (1 - \text{FPR})),
\end{equation}
\vskip -0.05in
\noindent where $\alpha$ and $\beta$ are weighting factors subject to $\alpha + \beta = 1$. In this formulation, the first component is Recall, which measures the proportion of correctly identified poisoned models among all actual poisoned models. The coefficient $\alpha$ explicitly controls the emphasis placed on this detection capability; thus, a larger $\alpha$ makes the evaluation favor defense methods that successfully identify a higher proportion of attackers. Conversely, FPR refers to the proportion of benign models that are incorrectly classified as poisoned (false alarms). The coefficient $\beta$ governs the weight of the $(1-\text{FPR})$ term, which represents the system's specificity. A larger $\beta$ prioritizes minimizing misclassification errors, favoring conservative defenses that preserve the participation of benign clients.

\begin{figure*}[t]
	\begin{center}
		\centerline{\includegraphics[width=\textwidth]{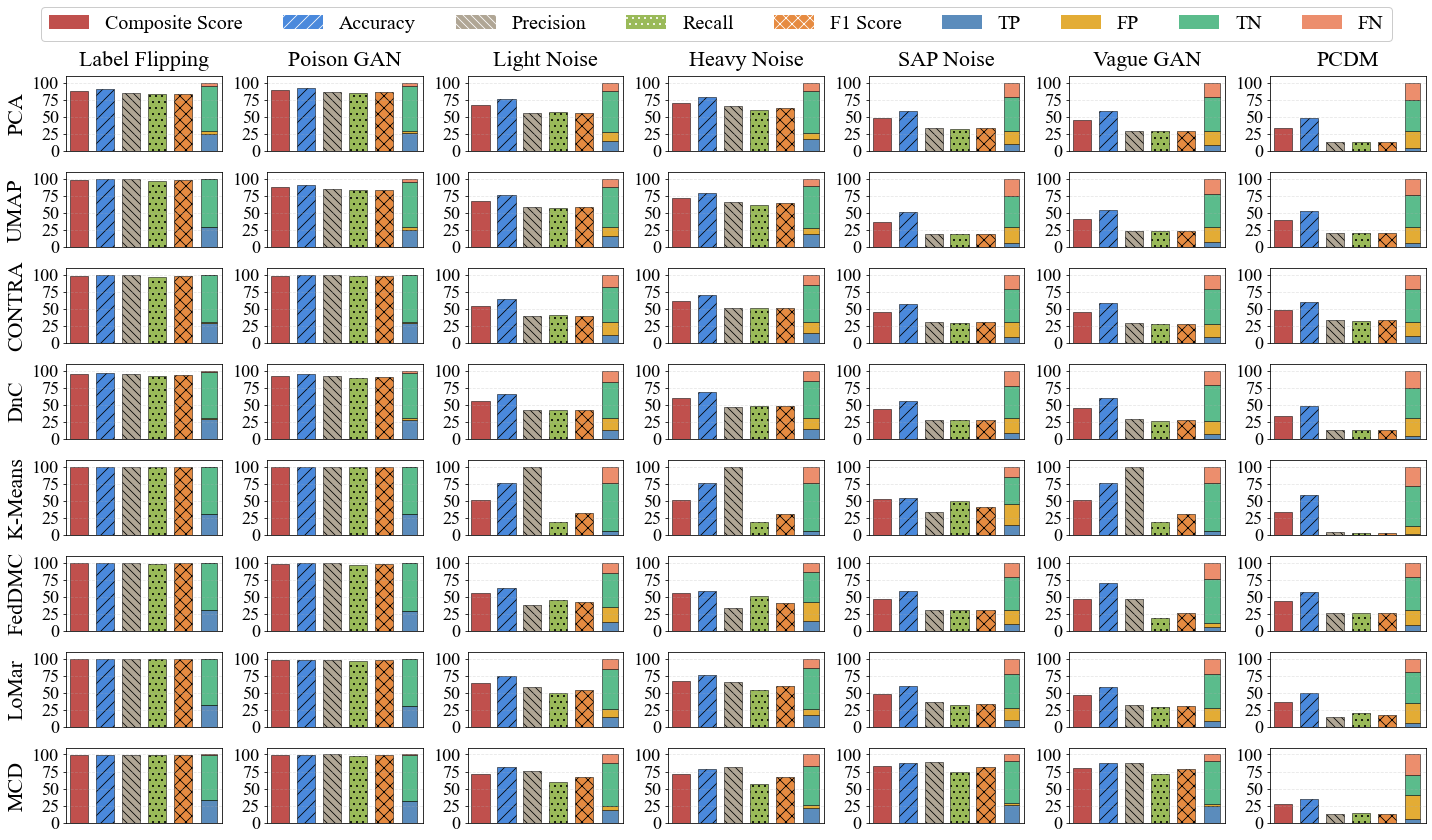}}
		\vskip -0.15in
		\caption{Defense methods performance breakdown by attack type.}
		\label{defense_all}
	\end{center}
    \vskip -0.45in
\end{figure*}

In this paper, we adopt a consistent configuration of $\alpha = 0.6$ and $\beta = 0.4$ across all datasets. This parameter choice reflects the asymmetric risk profile inherent in FL security: the cost of a "false negative" (missing a poisoning attack that corrupts the global model) is typically severer than that of a "false positive" (temporarily excluding a benign update). Consequently, by setting $\alpha > \beta$, we bias the composite score to slightly prioritize detection sensitivity to ensure system robustness, while still maintaining a strong constraint on false positive rates to avoid resource waste. This fixed configuration provides a standardized metric (ranging from 0 to 100) for security-oriented scenarios, enabling consistent comparison across different defenses.
%where $\alpha$ and $\beta$ are weighting factors such that $\alpha + \beta = 1$. In this formulation, recall measures the proportion of correctly identified poisoned models among all actual poisoned models, while FPR refers to the proportion of benign models that are incorrectly classified as poisoned. By incorporating both metrics, the composite score provides a holistic assessment of a defense's sensitivity and specificity. In our experiments, we set $\alpha = 0.6$ and $\beta = 0.4$, prioritizing detection sensitivity while still penalizing false positives. This normalized score (ranging from 0 to 100) allows consistent comparison across different defenses and attack types.

Fig.~\ref{defense_all} systematically presents a comparative evaluation of the performance of eight defense methods against seven types of model poisoning attacks on Fashion-MNIST dataset. Each subplot contains three components: (1) the rightmost stacked bar chart displays the raw data in terms of true positives (TP), false positives (FP), true negatives (TN), and false negatives (FN); (2) the middle four patterned bars show four standard evaluation metrics: accuracy, precision, recall, and F1-score; and (3) the leftmost solid bar represents the composite score, which jointly reflects a defense's sensitivity to poisoned models and its ability to avoid false positives.

Specifically, these metrics are calculated as follows:  
\begin{itemize}
    \item \textbf{Accuracy} = $\frac{TP + TN}{TP + TN + FP + FN}$, indicating the proportion of correctly identified models.

    \item \textbf{Precision} = $\frac{TP}{TP + FP}$, indicating the proportion of truly poisoned models among those classified as malicious.

    \item \textbf{Recall} = $\frac{TP}{TP + FN}$, indicating the proportion of poisoned models that are correctly identified as malicious.

    \item \textbf{F1-score} = $2 \times \frac{\text{Precision} \times \text{Recall}}{\text{Precision} + \text{Recall}}$, indicating the harmonic mean of precision and recall to balance both metrics.

\end{itemize}

Notably, existing defense methods perform well against traditional attacks such as label flipping, demonstrate moderate effectiveness against noise-based attacks, but struggle against generative poisoning methods. Among them, MCD performs relatively better against noise-based and GAN-based attacks, yet still fails to effectively detect our PCDM attack. Overall, the results clearly demonstrate that PCDM attack exhibits significantly enhanced stealth compared to other attack methods when confronted with various defense mechanisms. Its unique stealth attack mechanism makes it particularly challenging for existing defense systems to detect effectively.

\textbf{Performance Evaluation under Robust Aggregation Defenses.} 
To comprehensively evaluate the overall threat potential of the PCDM attack within robust FL environments, this study systematically assesses its performance against a variety of advanced robust aggregation defense mechanisms. The selected aggregation methods include: FedAvg (serving as the non-robust baseline), Multi-Krum, SignGuard, and the recently proposed LASA, which integrates pre-aggregation sparsification with layer-wise adaptive filtering. Experiments are conducted on the CIFAR-10 and CIFAR-100 datasets. To clearly demonstrate the attack effectiveness of PCDM, we adopt the superior ResNet-18 architecture as the backbone model, utilizing its robust feature extraction capabilities to establish a high-performance baseline. The federated training spans 200 rounds with varying proportions of malicious clients (0\%, 10\%, 20\%, and 30\%) subject to the PCDM attack. The final global model test accuracy serves as the primary metric, simultaneously quantifying both the destructiveness and stealthiness of the PCDM attack in a robust setting, thereby offering a comprehensive reflection of its overall performance.

% The main experimental results are summarized in Table~\ref{tab:robust_cifar10_100}. Overall, the PCDM attack poses a substantial threat to all tested robust defense methods. Even under the deployment of state-of-the-art defense mechanisms such as LASA, the PCDM attack still precipitates a significant degradation in global model accuracy. These results demonstrate that the highly stealthy poisoned data generated by PCDM exhibits strong adaptability. Despite being confronted with the most advanced adaptive robust defenses, PCDM effectively achieves poisoning through its fine-grained generative capability based on diffusion models, causing a noticeable decline in overall performance. This underscores the potent destructive potential of PCDM as a novel generative data poisoning attack.

\begin{table}[ht]
\vskip -0.15in
\centering
\caption{Global Model Accuracy (\%) under Various Robust Aggregation Defenses with PCDM Attack}
\label{tab:robust_cifar10_100}
\begin{tabular*}{\columnwidth}{@{\extracolsep{\fill}}llcccc@{}}
\toprule
\multirow{2}{*}{\textbf{Dataset}}   & \multirow{2}{*}{\textbf{Method}} & \multicolumn{4}{c}{\textbf{Malicious Clients Percentage $\alpha$}} \\ 
\cmidrule(l){3-6} 
                           &                         & 0\%          & 10\%         & 20\%         & 30\%         \\ \midrule
\multirow{5}{*}{CIFAR-10}  & FedAvg                  & 90.25        & 88.79        & 87.61        & 84.73        \\
                           
                           & Multi-Krum              & 83.51        & 83.4         & 83.11        & 82.79        \\
                           & SignGuard               & 89.9         & 87.69        & 87.3         & 85.32        \\
                           & LASA                    & 91.95        & 89.54        & 88.82        & 85.84        \\ 
% \hline
\midrule
\multirow{5}{*}{CIFAR-100} & FedAvg                  & 65.99        & 63.59        & 60.73        & 55.26        \\
                           
                           & Multi-Krum              & 52.66        & 52.38        & 51.69        & 50.68        \\
                           & SignGuard               & 63.64        & 61.73        & 60.89        & 56.83        \\
                           & LASA                    & 63.96        & 63.23        & 62.79        & 61.15        \\ \bottomrule
\end{tabular*}
\vskip -0.05in
\end{table}

The results in Table~\ref{tab:robust_cifar10_100} shows that the PCDM attack poses a substantial threat to all tested defenses. Even under state-of-the-art defenses like LASA, it still causes a significant degradation in global model accuracy. This demonstrates that the highly stealthy poisoned data generated by PCDM exhibits strong adaptability and can effectively bypass current advanced robust defenses, highlighting its potent destructive capability as a novel generative data poisoning attack.

\textbf{Applicability Verification in Wireless Environments.}
% To further validate the applicability of the proposed PCDM attack in real-world wireless network scenarios, we conduct additional experiments on the VRAI dataset~\cite{Wang2019vehicle}, which is a wireless-specific dataset collected from practical vehicular sensing environments. 
To further validate PCDM's applicability in real-world wireless scenarios, we conduct experiments on the VRAI dataset~\cite{Wang2019vehicle}, a wireless-specific dataset collected from practical vehicular sensing environments. VRAI contains diverse vehicle images captured under realistic wireless transmission conditions, making it a representative benchmark for assessing attack effectiveness beyond conventional image classification datasets. In our experiments, we utilize the training subset of VRAI consisting of 66,113 images, split 80\%/20\% for federated training and testing, focusing on the 7-class vehicle type classification task.
% and perform a stratified split, allocating 80\% of the data for federated training and the remaining 20\% for testing. We focus exclusively on the vehicle type classification task with 7 semantic categories.

\begin{figure}[h]
\vskip -0.1in
	\begin{center}
    \centerline{\includegraphics[width=\columnwidth]{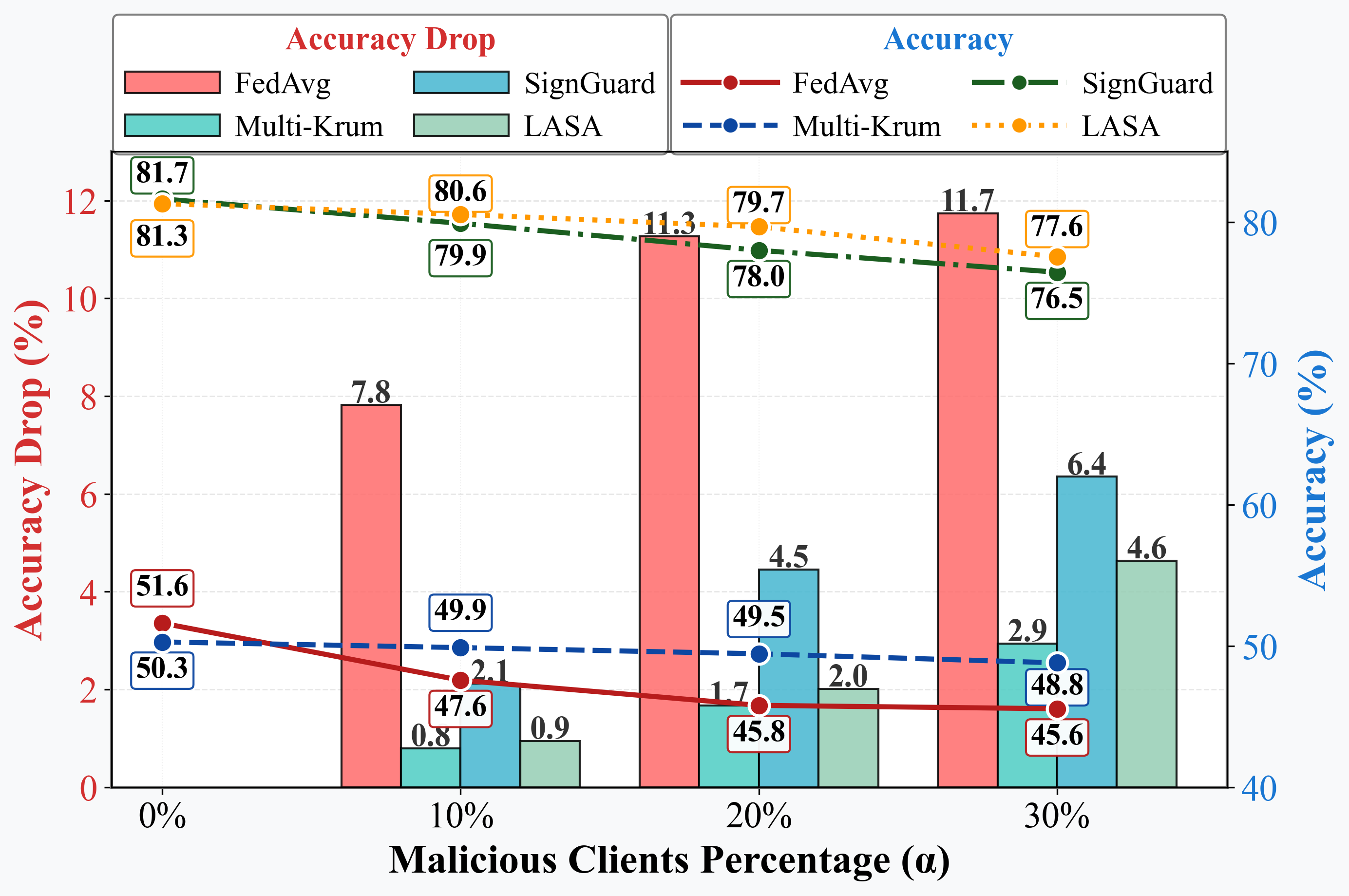}}
        \vskip -0.1in
		\caption{Accuracy of Different Defense Methods Against Various Malicious Client Ratios on VRAI Dataset.}
		\label{fig:vari}
	\end{center}
    \vskip -0.35in
\end{figure}

Fig.~\ref{fig:vari} illustrates 
% the evolution of global test accuracy and the corresponding accuracy degradation on the VRAI dataset as the proportion of malicious clients increases.
the global test accuracy and its degradation on VRAI as the malicious client ratio rises. 
Unlike standard vision benchmarks, the VRAI dataset exhibits more complex data characteristics arising from real-world wireless environments, including diverse viewpoints, environmental noise, and acquisition variability. Despite this, the PCDM still causes noticeable performance degradation across all evaluated aggregation strategies. 
% As the malicious client ratio grows from 10\% to 30\%, the global model accuracy shows a clear and steady decline, indicating that the attack effectiveness of PCDM remains stable and scalable even in realistic wireless FL systems.
The consistent accuracy decline as the malicious ratio increases from 10\% to 30\% indicates that the effectiveness of PCDM attack remains stable and scalable in realistic wireless FL systems.

Overall, the results show that the poisoned data generated by PCDM preserves its stealthiness and effectiveness in wireless-specific scenarios, where data distributions and system conditions are inherently more challenging. 
Even against advanced robust aggregation mechanisms, PCDM continues to undermine learning without triggering obvious defensive responses. 
% Even when combined with advanced robust aggregation mechanisms, PCDM continues to undermine the learning process without triggering obvious defensive responses. 
This confirms that PCDM is not limited to controlled benchmark settings but exhibits strong generalizability and practical threat potential in real-world wireless FL deployments.

\textbf{Statistical Indistinguishability Verification.}
Complementing the qualitative visualizations presented earlier, the successful evasion of the aforementioned eleven defense and aggregation mechanisms provides a rigorous, multi-dimensional validation of PCDM's invisibility. As summarized in Table~\ref{tab:stealth_metrics_v2}, these defenses define normality through strict statistical metrics ranging from Euclidean norms and clustering scores to cosine similarity and distributional divergence. Consequently, PCDM's consistent ability to circumvent these filters demonstrates that its generated poisonous updates maintain statistically insignificant deviations from benign updates across diverse metric spaces. This empirically confirms the attack's superior evasiveness regarding critical quantitative indicators, such as perturbation magnitude, directional alignment, and distributional consistency, thereby satisfying the stringent requirements for robust evaluation.

\begin{table}[ht]
\vskip -0.15in
\centering
\caption{Quantitative Stealthiness Validation: Mapping Deployed Defenses to Statistical Metrics.}
\label{tab:stealth_metrics_v2}
\renewcommand{\arraystretch}{1.4} % 三线表通常稍微把行高调大一点点，更美观
\footnotesize 
\setlength{\tabcolsep}{4pt} % 去掉竖线后，可以稍微把每一列的内部间距加大一点

% 定义列宽，去掉竖线 | 
\begin{tabular}{p{0.13\columnwidth} p{0.34\columnwidth} p{0.45\columnwidth}}
\toprule
\textbf{Type} & \textbf{Quantitative Metric} & \textbf{Validated Baselines} \\ 
\midrule % 标题下的分割线

% 第一组：Distance
\multirow{3}{=}{\textbf{Distance}} 
 & Euclidean Summation & Multi-Krum \cite{blanchard2017machine} \\
 % 为了美观，这里用 addlinespace 增加一点组内间距，或者直接换行
 \addlinespace[2pt]
 & Subspace Outlier Score & PCA \cite{tolpegin2020data}, K-Means \cite{li2022robust}, UMAP \cite{upreti2022defending}, FedDMC \cite{mu2024feddmc} \\ 
\midrule % 类别间分割线

% 第二组：Direction
\multirow{2}{=}{\textbf{Direction}} 
 & Cosine Similarity & CONTRA \cite{awan2021contra} \\
 \addlinespace[2pt]
 & Projection \& Signal Stats & DnC \cite{shejwalkar2021manipulating}, SignGuard \cite{xu2022byzantine} \\ 
\midrule

% 第三组：Distribution
\multirow{2}{=}{\textbf{Distrib.}} 
 & Kernel Density & LoMar \cite{li2021lomar} \\
 & Parameter Consistency & MCD \cite{sun2024gan} \\ 
\midrule

% 第四组：Hybrid
\textbf{Hybrid} & Magnitude \& Direction & LASA \cite{xu2025achieving} \\ 
\bottomrule % 底部粗线

\end{tabular}
\vskip -0.25in
\end{table}

\subsection{Ablation Study}
To comprehensively assess the effectiveness of PCDM in the context of data poisoning attacks, a series of experiments were conducted across multiple datasets focusing on three principal aspects: the distinctive characteristics of the poisoned data, the effectiveness of the poisoning attacks, and the associated computational overhead. The evaluation involved comparative analyses among three models: DDPM, PCDM without the poison vector $\mathbf{v}$, and the complete PCDM framework, where "w/o $\mathbf{v}$" refers to the exclusion of the poisoning vector.

\begin{figure}[h]
\vskip -0.1in
	\begin{center}
        \centerline{\includegraphics[width=\columnwidth]{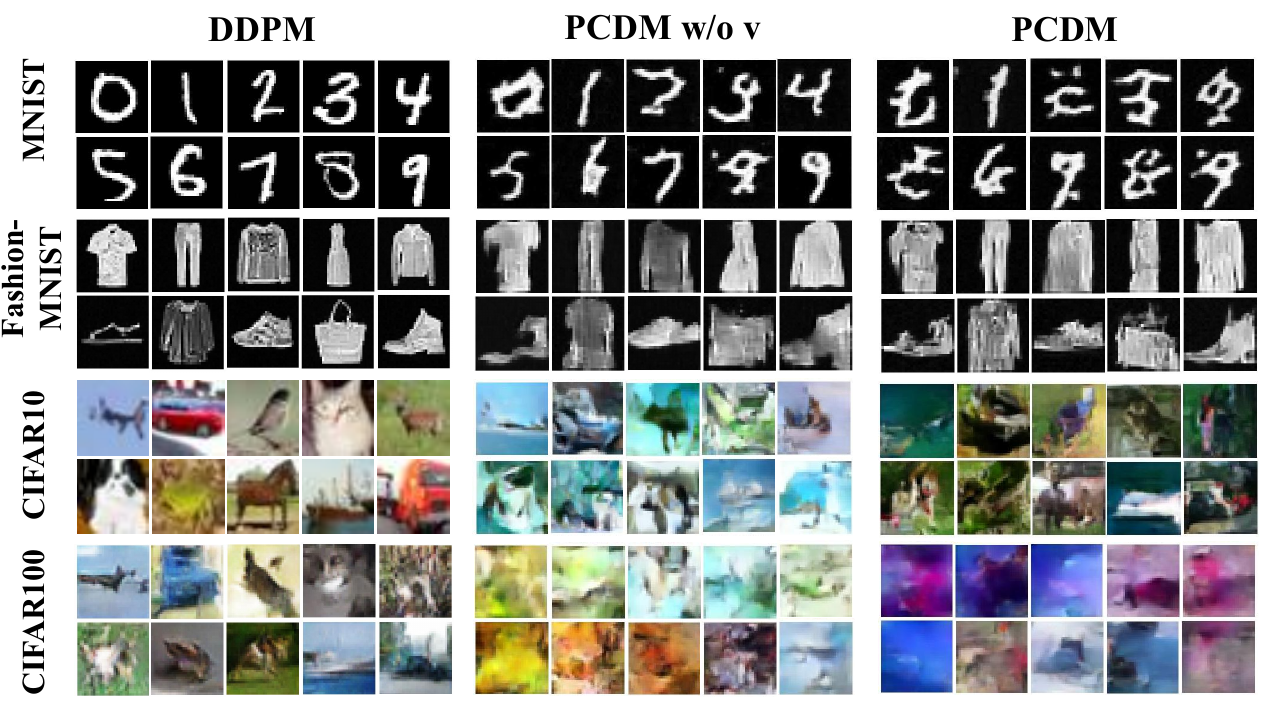}}
        \vskip -0.1in
		\caption{Ablation study comparison of poisoned data.}
		\label{fig:Ablation}
	\end{center}
	\vskip -0.3in
\end{figure}

As depicted in Fig.~\ref{fig:Ablation}, PCDM, after capturing the salient features of the training data, is capable of synthesizing poisoned samples that are noisy yet retain the authentic attributes of real data. This generative behavior is markedly different from that observed with DDPM. Furthermore, the integration of the poison vector within PCDM enables the production of poisoned data with substantially enhanced attack characteristics, demonstrating the indispensability of the poison vector $\mathbf{v}$ for executing efficient data poisoning attacks.

\begin{figure}[h]
\vskip -0.1in
	\begin{center}
        \centerline{\includegraphics[width=\columnwidth]{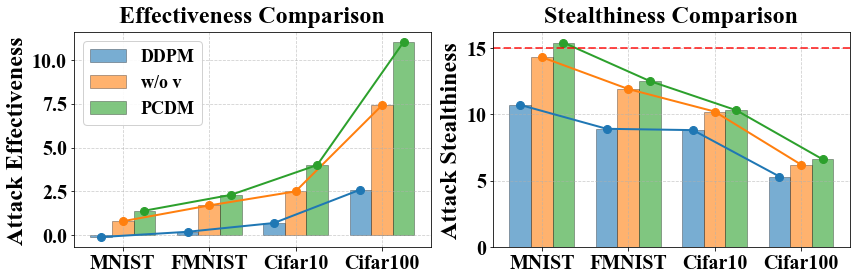}}
        \vskip -0.1in
		\caption{Ablation study comparison of attack effectiveness and stealthiness.}
		\label{fig:Ablation2}
	\end{center}
	\vskip -0.3in
\end{figure}

The quantitative results, as shown in Fig.~\ref{fig:Ablation2}, indicate that the data synthesized by DDPM lack effective attack capabilities. For complex datasets such as CIFAR-100, any apparent attack effects with DDPM are primarily attributable to the restricted capacity resulting from limited client-side data rather than genuine adversarial traits. Crucially, our analysis reveals that the inclusion of the poisoning vector $\mathbf{v}$ yields a systematic improvement in both stealthiness and aggressiveness compared to the baseline and the version without $\mathbf{v}$. While this performance gain is universal across all tested scenarios, the magnitude of the benefit correlates with dataset complexity. Specifically, the improvements are significantly more pronounced on high-dimensional, complex datasets (e.g., CIFAR-100 and CIFAR-10) than on simpler ones (e.g., MNIST and Fashion-MNIST).

With respect to stealthiness, experiments reveal that when the model distance metric remains below approximately 15 (as indicated by the dashed line), the attack is sufficiently covert to evade most existing detection mechanisms. Although the introduction of the poison vector leads to a variation in model updates, the stealthiness metric remains well within acceptable bounds (below the threshold). This confirms that PCDM successfully optimizes the trade-off between invisibility and utility. Collectively, these findings underscore the necessity of incorporating the poison vector $\mathbf{v}$ in PCDM to achieve a potent attack, particularly in complex data environments.
%The results of the ablation studies, as shown in Fig.~\ref{fig:Ablation2}, indicate that the data synthesized by DDPM lack effective attack capabilities. For complex datasets such as CIFAR100, any apparent attack effects observed with DDPM are primarily attributable to the restricted capacity resulting from limited client-side data, rather than genuine adversarial traits. With respect to stealthiness, experimental results reveal that when the model distance metric remains below approximately 15 (as indicated by the dashed line), the attack is sufficiently covert to evade most existing detection mechanisms. This suggests a minimal difference between poisoned and benign models, ensuring a satisfactory level of stealthiness. While the introduction of the poison vector in PCDM leads to a minor decrease in stealthiness, the reduction is within acceptable bounds. This trade-off results in a significant boost in attack effectiveness, yielding more potent and threatening data poisoning attacks. Collectively, these findings underscore the necessity of incorporating the poison vector in PCDM to achieve both effective and stealthy poisoning attacks.

\begin{table}[h!]
\vskip -0.15in
    \centering
    \caption{Comparison of Time Overhead for Poisoned Data Generation (s)}
    \label{tab:overhead}
    \begin{tabular*}{\columnwidth}{@{\extracolsep{\fill}}lcccc@{}}
        \toprule
        \textbf{Model} & \textbf{MNIST} & \textbf{Fashion-MNIST} & \textbf{CIFAR10} & \textbf{CIFAR100} \\ 
        \midrule
        DDPM & 1126.23 & 1114.10 & 1445.37 & 1462.69 \\
        PCDM w/o \textbf{v} & 167.32 & 166.97 & 196.41 & 197.20 \\
        PCDM & 174.28 & 175.83 & 211.01 & 215.08 \\
        \bottomrule
    \end{tabular*}
    \vskip -0.05in
\end{table}

Additionally, the computational time required for training and generating an equivalent amount of poisoned data (3{,}000 samples) using the three models on client devices equipped with an NVIDIA A100-40GB GPU is summarized in Table~\ref{tab:overhead}. It is evident that PCDM incurs considerably less overhead for poisoned data generation compared to DDPM, and the inclusion of the poison vector introduces negligible extra computational cost. 

In summary, PCDM offers an efficient, stealthy, and lightweight data poisoning attack strategy. Owing to these advantages, PCDM poses a practical and formidable threat to FL systems, thereby highlighting the urgent need for robust defense mechanisms in real-world FL deployments.

\subsection{Efficiency Analysis and Resource Consumption}

To validate the deployment feasibility on resource-constrained edge devices, we conducted an efficiency benchmark measuring runtime, peak memory usage, and FLOPs on an NVIDIA A100 GPU against DDPM and VagueGAN baselines. As detailed in Table~\ref{tab:overhead}, PCDM fundamentally overcomes the computational bottleneck of standard diffusion models. By employing the jumping diffusion strategy, significantly reducing the Markov chain length, PCDM achieves a drastic 85\% reduction in latency and computational cost compared to DDPM (e.g., reducing CIFAR-10 FLOPs from 45k G to 6.7k G). This improvement confirms that our sparse sampling strategy effectively eliminates the heavy overhead typically associated with diffusion-based generation.

\begin{table}[h]
\vskip -0.1in
    \centering
    \caption{Efficiency Benchmark: Runtime (s), FLOPs (G), and Peak Memory (MB) for Generating 3,000 Samples.}
    \label{tab:overhead}
    \resizebox{\columnwidth}{!}{%
    \begin{tabular}{l|l|c c c}
        \toprule
        \textbf{Dataset} & \textbf{Model} & \textbf{Time (s)} & \textbf{FLOPs (G)} & \textbf{Mem (MB)} \\ 
        \midrule
        \multirow{3}{*}{MNIST} 
        & VagueGAN & 155.60 & 4450.2 & 1850 \\
        & DDPM                       & 1126.23 & 32450.6 & 2240 \\
        & \textbf{PCDM}       & \textbf{174.28} & \textbf{5108.3} & \textbf{2255} \\
        \midrule
        \multirow{3}{*}{\shortstack{Fashion-\\MNIST}} 
        & VagueGAN                   & 158.45 & 4510.5 & 1880 \\
        & DDPM                       & 1114.10 & 32105.2 & 2240 \\
        & \textbf{PCDM}       & \textbf{175.83} & \textbf{5152.7} & \textbf{2258} \\
        \midrule
        \multirow{3}{*}{CIFAR-10} 
        & VagueGAN                   & 189.20 & 5980.4 & 2950 \\
        & DDPM                       & 1445.37 & 45220.9 & 3420 \\
        & \textbf{PCDM}       & \textbf{211.01} & \textbf{6760.5} & \textbf{3445} \\
        \midrule
        \multirow{3}{*}{CIFAR-100} 
        & VagueGAN                   & 194.50 & 6120.1 & 2980 \\
        & DDPM                       & 1462.69 & 45780.4 & 3420 \\
        & \textbf{PCDM}       & \textbf{215.08} & \textbf{6890.2} & \textbf{3448} \\
        \bottomrule
    \end{tabular}%
    }
    \vskip -0.1in
\end{table}
\label{sec:efficiency}

Furthermore, PCDM effectively closes the efficiency gap between diffusion and GAN-based architectures. despite GANs' inherent single-step speed advantage, our results indicate that PCDM is highly competitive with the state-of-the-art VagueGAN, incurring only a marginal runtime increase (10–15\%) and comparable memory usage. This demonstrates that the optimized diffusion process achieves efficiency levels practically indistinguishable from GANs, rendering it suitable for real-time wireless scenarios.

Beyond quantitative metrics, PCDM ensures real-world viability through flexible deployment strategies. For extremely resource-limited devices, an attacker can adopt a \textit{train-then-distribute} strategy~\cite{sun2024gan}, where a powerful server pre-trains the model and distributes lightweight weights for low-cost client-side inference. Alternatively, the framework supports heterogeneous deployment, allowing clients to adaptively select sampling hyperparameters based on local hardware states, ensuring robust attack execution across diverse edge environments.

\subsection{Performance Trade-off Analysis}

The optimization formulation in Eq.~(46) highlights the inherent tension between poisoning intensity and detectability. While enhancing poisoning intensity improves attack effectiveness, it inevitably increases exposure to defenses. PCDM balances these conflicting objectives through coordinated hyperparameter tuning.

We verify this trade-off using Fashion-MNIST, analyzing the impact of the noise vector $\mathbf{v}$ and diffusion parameters. Following Sec.~\ref{stea}, we quantify stealthiness using the Euclidean distance of PCA-reduced model updates.

\begin{figure}[H]
	\vskip -0.2in
	\begin{center}
		\centerline{\includegraphics[width=\columnwidth]{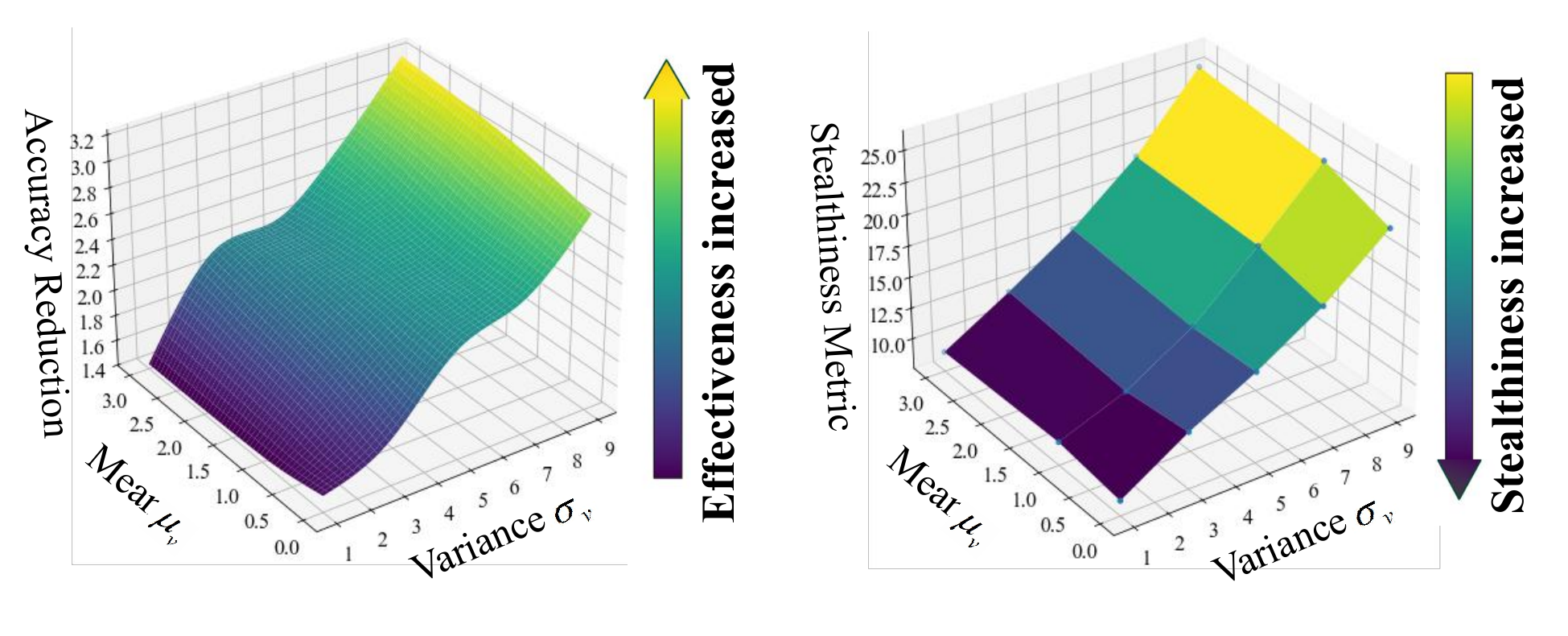}}
		\vskip -0.25in
		\caption{Sensitivity analysis of attack performance with respect to poisoning vector noise parameters.}
		\label{fig:last}
	\end{center}
	\vskip -0.2in
\end{figure}

\begin{figure}[H]
	\vskip -0.15in
	\begin{center}
		\centerline{\includegraphics[width=\columnwidth]{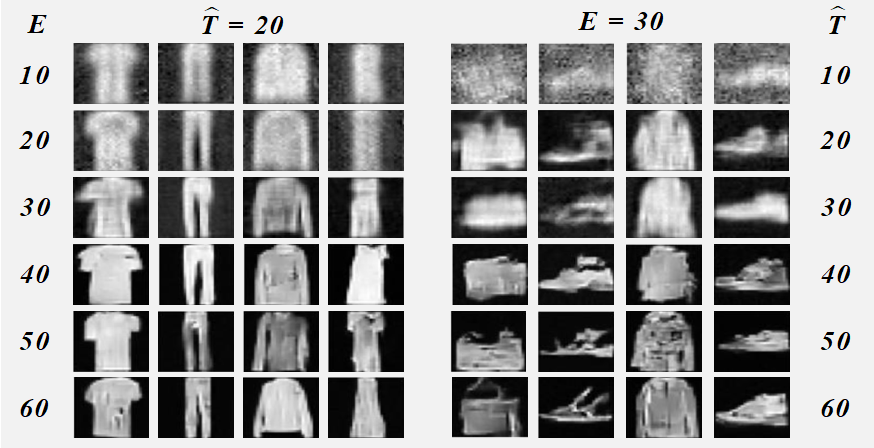}}
		\vskip -0.15in
		\caption{Impact of $E$ and $\hat{T}$ on the effectiveness-stealthiness trade-off.}
		\label{fig:EandT}
	\end{center}
	\vskip -0.25in
\end{figure}

As shown in Fig.~\ref{fig:last}, higher noise levels in $\mathbf{v}$ positively correlate with attack effectiveness (accuracy reduction) but compromise stealthiness (increased model distance). To strike an optimal balance, we recommend moderate settings of $\sigma_v \approx 1$ and $\mu_v \approx 5$. Under these conditions, the model distance remains below 15, which is a threshold ensuring sufficient stealthiness, while maintaining substantial accuracy degradation. Beyond noise parameters, the diffusion hyperparameters, specifically the training epochs $E$ and the total diffusion steps $\hat{T}$, critically govern sample fidelity and the resulting trade-off. We investigate their influence in Fig.~\ref{fig:EandT}.

Results indicate that increasing $E$ or $\hat{T}$ enhances the fidelity of poisoned samples, aligning them closer to benign distributions. This improves stealthiness significantly but mitigates the aggressiveness of the attack. Unlike $\sigma_v$ and $\mu_v$, which have consistent effects, optimal $E$ and $\hat{T}$ are sensitive to dataset complexity. Complex or high-resolution data typically demand higher values to ensure fidelity.

Based on our analysis, we adopt $E=30$ and $\hat{T}=20$ for most experiments of this paper. This configuration offers a robust baseline, producing sufficiently realistic samples to evade detection while retaining high attack potency. Practitioners should adjust these values according to the target data complexity to achieve an optimal trade-off.

\section{Conclusion and Future Work}

In this paper, we have proposed PCDM, a lightweight and highly stealthy data poisoning attack for FL systems, which introduced a specialized poisoning vector and an innovative jumping diffusion strategy to enable highly stealthy and efficient attacks. We have also provided a comprehensive theoretical analysis demonstrating PCDM's capability to control poisoned data, thereby offering guidance for balancing attack effectiveness and stealthiness. In addition, a large-scale comparative evaluation has been carried out on five datasets including the real-world wireless-specific VRAI benchmark, involving eleven diverse defense mechanisms covering advanced robust aggregation strategies and seven representative poisoning attacks, which is unprecedented in FL poisoning research. Extensive experimental results have demonstrated the superior performance of PCDM attacks, which poses a significant risk to the security of FL systems. In the future, we intend to further investigate countermeasures against such generative AI-based stealthy attacks.

\bibliography{bio2}

@Inproceedings{bagdasaryan2020backdoor,
	title="How to backdoor federated learning",
	author="Bagdasaryan, Eugene and Veit, Andreas and Hua, Yiqing and Estrin, Deborah and Shmatikov, Vitaly",
	booktitle="International conference on artificial intelligence and statistics",
	pages="2938--2948",
	year="2020",
	organization="PMLR"
}

@article{zhang2020poisongan,
	title="PoisonGAN: Generative poisoning attacks against federated learning in edge computing systems",
	author="Zhang, Jiale and Chen, Bing and Cheng, Xiang and Binh, Huynh Thi Thanh and Yu, Shui",
	journal="IEEE Internet of Things Journal",
	volume="8",
	number="5",
	pages="3310--3322",
	year="2020",
	publisher="IEEE"
}

@inproceedings{tolpegin2020data,
	title="Data poisoning attacks against federated learning systems",
	author="Tolpegin, Vale and Truex, Stacey and Gursoy, Mehmet Emre and Liu, Ling",
	booktitle="Computer Security--ESORICS 2020: 25th European Symposium on Research in Computer Security, ESORICS 2020, Guildford, UK, September 14--18, 2020, Proceedings, Part I 25",
	pages="480--501",
	year="2020",
	organization="Springer"
}

@inproceedings{shejwalkar2021manipulating,
	title="Manipulating the byzantine: Optimizing model poisoning attacks and defenses for federated learning",
	author="Shejwalkar, Virat and Houmansadr, Amir",
	booktitle="NDSS",
	year="2021"
}

@inproceedings{awan2021contra,
	title="Contra: Defending against poisoning attacks in federated learning",
	author="Awan, Sana and Luo, Bo and Li, Fengjun",
	booktitle="Computer Security--ESORICS 2021: 26th European Symposium on Research in Computer Security, Darmstadt, Germany, October 4--8, 2021, Proceedings, Part I 26",
	pages="455--475",
	year="2021",
	organization="Springer"
}

@article{li2021lomar,
	title="Lomar: A local defense against poisoning attack on federated learning",
	author="Li, Xingyu and Qu, Zhe and Zhao, Shangqing and Tang, Bo and Lu, Zhuo and Liu, Yao",
	journal="IEEE Transactions on Dependable and Secure Computing",
	year="2021",
	publisher="IEEE"
}

@inproceedings{mcmahan2017communication,
	title="Communication-efficient learning of deep networks from decentralized data",
	author="McMahan, Brendan and Moore, Eider and Ramage, Daniel and Hampson, Seth and y Arcas, Blaise Aguera",
	booktitle="Artificial intelligence and statistics",
	pages="1273--1282",
	year="2017",
	organization="PMLR"
}

@INPROCEEDINGS{10287523,
	author="Sun, Wei and Gao, Bo and Xiong, Ke and Lu, Yang and Wang, Yuwei",
	booktitle="2023 20th Annual IEEE International Conference on Sensing, Communication, and Networking (SECON)", 
	title="VagueGAN: A GAN-Based Data Poisoning Attack Against Federated Learning Systems", 
	year="2023",
	volume="",
	number="",
	pages="321-329",
	doi="10.1109/SECON58729.2023.10287523"}

@article{rodriguez2023survey,
	title="Survey on federated learning threats: Concepts, taxonomy on attacks and defences, experimental study and challenges",
	author="Rodr{\'\i}guez-Barroso, Nuria and Jim{\'e}nez-L{\'o}pez, Daniel and Luz{\'o}n, M Victoria and Herrera, Francisco and Mart{\'\i}nez-C{\'a}mara, Eugenio",
	journal="Information Fusion",
	volume="90",
	pages="148--173",
	year="2023",
	publisher="Elsevier"
}

@article{KASYAP2024122210,
	title = "Privacy-preserving and Byzantine-robust Federated Learning Framework using Permissioned Blockchain",
	author = "Harsh Kasyap and Somanath Tripathy",
	journal = "Expert Systems with Applications",
	volume = "238",
	pages = "122210",
	year = "2024",
}

@article{zhang2024visualizing,
	title={Visualizing the Shadows: Unveiling Data Poisoning Behaviors in Federated Learning},
	author={Zhang, Xueqing and Zhang, Junkai and Chow, Ka-Ho and Chen, Juntao and Mao, Ying and Rahouti, Mohamed and Li, Xiang and Liu, Yuchen and Wei, Wenqi},
	journal={arXiv preprint arXiv:2405.16707},
	year={2024}
}

@article{wei2023demystifying,
	title={Demystifying data poisoning attacks in distributed learning as a service},
	author={Wei, Wenqi and Chow, Ka-Ho and Wu, Yanzhao and Liu, Ling},
	journal={IEEE Transactions on Services Computing},
	year={2023},
	publisher={IEEE}
}

@article{ho2020denoising,
	title="Denoising diffusion probabilistic models",
	author="Ho, Jonathan and Jain, Ajay and Abbeel, Pieter",
	journal="Advances in neural information processing systems",
	volume="33",
	pages="6840--6851",
	year="2020"
}

@article{sun2024gan,
	title={A GAN-Based Data Poisoning Attack Against Federated Learning Systems and Its Countermeasure},
	author={Sun, Wei and Gao, Bo and Xiong, Ke and Wang, Yuwei and Fan, Pingyi and Letaief, Khaled Ben},
	journal={arXiv preprint arXiv:2405.11440},
	year={2024}
}

@article{deng2012mnist,
	title={The mnist database of handwritten digit images for machine learning research [best of the web]},
	author={Deng, Li},
	journal={IEEE signal processing magazine},
	volume={29},
	number={6},
	pages={141--142},
	year={2012},
	publisher={IEEE}
}

@article{xiao2017fashion,
	title={Fashion-mnist: a novel image dataset for benchmarking machine learning algorithms},
	author={Xiao, Han and Rasul, Kashif and Vollgraf, Roland},
	journal={arXiv preprint arXiv:1708.07747},
	year={2017}
}

@article{krizhevsky2009learning,
	title={Learning multiple layers of features from tiny images},
	author={Krizhevsky, Alex and Hinton, Geoffrey and others},
	year={2009},
	publisher={Toronto, ON, Canada}
}

@inproceedings{gragnaniello2018analysis,
	title={Analysis of adversarial attacks against CNN-based image forgery detectors},
	author={Gragnaniello, Diego and Marra, Francesco and Poggi, Giovanni and Verdoliva, Luisa},
	booktitle={2018 26th European Signal Processing Conference (EUSIPCO)},
	pages={967--971},
	year={2018},
	organization={IEEE}
}

@article{upreti2022defending,
	title={Defending against Label-Flipping Attacks in Federated Learning Systems with UMAP},
	author={Upreti, Deepak and Kim, Hyunil and Yang, Eunmok and Seo, Changho},
	year={2022}
}

@inproceedings{shen2016auror,
	title={Auror: Defending against poisoning attacks in collaborative deep learning systems},
	author={Shen, Shiqi and Tople, Shruti and Saxena, Prateek},
	booktitle={Proceedings of the 32nd annual conference on computer security applications},
	pages={508--519},
	year={2016}
}

@article{yang2023clean,
	title={Clean-label poisoning attacks on federated learning for IoT},
	author={Yang, Jie and Zheng, Jun and Baker, Thar and Tang, Shuai and Tan, Yu-an and Zhang, Quanxin},
	journal={Expert Systems},
	volume={40},
	number={5},
	pages={e13161},
	year={2023},
	publisher={Wiley Online Library}
}

@article{marfoq2021federated,
	title={Federated multi-task learning under a mixture of distributions},
	author={Marfoq, Othmane and Neglia, Giovanni and Bellet, Aur{\'e}lien and Kameni, Laetitia and Vidal, Richard},
	journal={Advances in Neural Information Processing Systems},
	volume={34},
	pages={15434--15447},
	year={2021}
}

@article{li2020review,
	title={A review of applications in federated learning},
	author={Li, Li and Fan, Yuxi and Tse, Mike and Lin, Kuo-Yi},
	journal={Computers \& Industrial Engineering},
	volume={149},
	pages={106854},
	year={2020},
	publisher={Elsevier}
}

@article{liu2024vertical,
	title={Vertical federated learning: Concepts, advances, and challenges},
	author={Liu, Yang and Kang, Yan and Zou, Tianyuan and Pu, Yanhong and He, Yuanqin and Ye, Xiaozhou and Ouyang, Ye and Zhang, Ya-Qin and Yang, Qiang},
	journal={IEEE Transactions on Knowledge and Data Engineering},
	year={2024},
	publisher={IEEE}
}

@article{nowroozi2025federated,
	title={Federated learning under attack: Exposing vulnerabilities through data poisoning attacks in computer networks},
	author={Nowroozi, Ehsan and Haider, Imran and Taheri, Rahim and Conti, Mauro},
	journal={IEEE Transactions on Network and Service Management},
	year={2025},
	publisher={IEEE}
}

@article{kasyap2024beyond,
	title={Beyond data poisoning in federated learning},
	author={Kasyap, Harsh and Tripathy, Somanath},
	journal={Expert Systems with Applications},
	volume={235},
	pages={121192},
	year={2024},
	publisher={Elsevier}
}

@article{wan2024data,
	title={Data and model poisoning backdoor attacks on wireless federated learning, and the defense mechanisms: A comprehensive survey},
	author={Wan, Yichen and Qu, Youyang and Ni, Wei and Xiang, Yong and Gao, Longxiang and Hossain, Ekram},
	journal={IEEE Communications Surveys \& Tutorials},
	year={2024},
	publisher={IEEE}
}

@article{borji2022pros,
	title={Pros and cons of GAN evaluation measures: New developments},
	author={Borji, Ali},
	journal={Computer Vision and Image Understanding},
	volume={215},
	pages={103329},
	year={2022},
	publisher={Elsevier}
}

@article{wang2023gan,
	title={Gan-generated faces detection: A survey and new perspectives},
	author={Wang, Xin and Guo, Hui and Hu, Shu and Chang, Ming-Ching and Lyu, Siwei},
	journal={ECAI 2023},
	pages={2533--2542},
	year={2023},
	publisher={IOS Press}
}

@incollection{wei2024data,
	title={Data Poisoning and Leakage Analysis in Federated Learning},
	author={Wei, Wenqi and Huang, Tiansheng and Yahn, Zachary and Singhal, Anoop and Loper, Margaret and Liu, Ling},
	booktitle={Handbook of Trustworthy Federated Learning},
	pages={73--108},
	year={2024},
	publisher={Springer}
}

@article{wen2023survey,
	title={A survey on federated learning: challenges and applications},
	author={Wen, Jie and Zhang, Zhixia and Lan, Yang and Cui, Zhihua and Cai, Jianghui and Zhang, Wensheng},
	journal={International Journal of Machine Learning and Cybernetics},
	volume={14},
	number={2},
	pages={513--535},
	year={2023},
	publisher={Springer}
}

@article{croitoru2023diffusion,
	title={Diffusion models in vision: A survey},
	author={Croitoru, Florinel-Alin and Hondru, Vlad and Ionescu, Radu Tudor and Shah, Mubarak},
	journal={IEEE Transactions on Pattern Analysis and Machine Intelligence},
	volume={45},
	number={9},
	pages={10850--10869},
	year={2023},
	publisher={IEEE}
}

@article{yang2023diffusion,
	title={Diffusion models: A comprehensive survey of methods and applications},
	author={Yang, Ling and Zhang, Zhilong and Song, Yang and Hong, Shenda and Xu, Runsheng and Zhao, Yue and Zhang, Wentao and Cui, Bin and Yang, Ming-Hsuan},
	journal={ACM Computing Surveys},
	volume={56},
	number={4},
	pages={1--39},
	year={2023},
	publisher={ACM New York, NY, USA}
}

@article{liu2024recent,
	title={Recent advances on federated learning: A systematic survey},
	author={Liu, Bingyan and Lv, Nuoyan and Guo, Yuanchun and Li, Yawen},
	journal={Neurocomputing},
	pages={128019},
	year={2024},
	publisher={Elsevier}
}

@article{li2024threats,
	title={Threats and Defenses in Federated Learning Life Cycle: A Comprehensive Survey and Challenges},
	author={Li, Yanli and Guo, Zhongliang and Yang, Nan and Chen, Huaming and Yuan, Dong and Ding, Weiping},
	journal={arXiv preprint arXiv:2407.06754},
	year={2024}
}

@article{cao2024survey,
	title={A survey on generative diffusion models},
	author={Cao, Hanqun and Tan, Cheng and Gao, Zhangyang and Xu, Yilun and Chen, Guangyong and Heng, Pheng-Ann and Li, Stan Z},
	journal={IEEE Transactions on Knowledge and Data Engineering},
	year={2024},
	publisher={IEEE}
}

@article{regenwetter2022deep,
	title={Deep generative models in engineering design: A review},
	author={Regenwetter, Lyle and Nobari, Amin Heyrani and Ahmed, Faez},
	journal={Journal of Mechanical Design},
	volume={144},
	number={7},
	pages={071704},
	year={2022},
	publisher={American Society of Mechanical Engineers}
}

@article{zhang2024fltracer,
	title={Fltracer: Accurate poisoning attack provenance in federated learning},
	author={Zhang, Xinyu and Liu, Qingyu and Ba, Zhongjie and Hong, Yuan and Zheng, Tianhang and Lin, Feng and Lu, Li and Ren, Kui},
	journal={IEEE Transactions on Information Forensics and Security},
	year={2024},
	publisher={IEEE}
}

@inproceedings{chen2024exploring,
	title={Exploring representational similarity analysis to protect federated learning from data poisoning},
	author={Chen, Gengxiang and Li, Kai and Abdelmoniem, Ahmed M and You, Linlin},
	booktitle={Companion Proceedings of the ACM on Web Conference 2024},
	pages={525--528},
	year={2024}
}

@article{wu2024challenges,
	title={Challenges and Countermeasures of Federated Learning Data Poisoning Attack Situation Prediction},
	author={Wu, Jianping and Jin, Jiahe and Wu, Chunming},
	journal={Mathematics},
	volume={12},
	number={6},
	pages={901},
	year={2024},
	publisher={MDPI}
}

@article{bengesi2024advancements,
	title={Advancements in Generative AI: A Comprehensive Review of GANs, GPT, Autoencoders, Diffusion Model, and Transformers.},
	author={Bengesi, Staphord and El-Sayed, Hoda and Sarker, Md Kamruzzaman and Houkpati, Yao and Irungu, John and Oladunni, Timothy},
	journal={IEEE Access},
	year={2024},
	publisher={IEEE}
}

@article{onsu2023cope,
  title={How to cope with malicious federated learning clients: an unsupervised learning-based approach},
  author={Onsu, Murat Arda and Kantarci, Burak and Boukerche, Azzedine},
  journal={Computer Networks},
  volume={234},
  pages={109938},
  year={2023},
  publisher={Elsevier}
}

@inproceedings{li2022robust,
  title={Robust federated learning based on metrics learning and unsupervised clustering for malicious data detection},
  author={Li, Jiaming and Zhang, Xinyue and Zhao, Liang},
  booktitle={Proceedings of the 2022 ACM Southeast Conference},
  pages={238--242},
  year={2022}
}

@article{mu2024feddmc,
  title={Feddmc: Efficient and robust federated learning via detecting malicious clients},
  author={Mu, Xutong and Cheng, Ke and Shen, Yulong and Li, Xiaoxiao and Chang, Zhao and Zhang, Tao and Ma, Xindi},
  journal={IEEE Transactions on Dependable and Secure Computing},
  year={2024},
  publisher={IEEE}
}

@inproceedings{Wang2019vehicle,
  title={Vehicle Re-identification in Aerial Imagery : Dataset and Approach},
  author={Peng, Wang and Bingliang, Jiao and Lu, Yang and Shizhou, Zhang and Wei, Wei and Yanning, Zhang},
  booktitle={Proc. IEEE Int. Conf. Comp. Vis.},
  year={2019}
}

@article{yazdinejad2025explainable,
  title={An Explainable and Privacy-Preserving Federated Learning Model for Threat Detection in Cyber-Physical-Social Systems},
  author={Yazdinejad, Abbas and Mohammadabadi, Zahra Dehghani and Dehghantanha, Ali and Srivastava, Gautam},
  journal={IEEE Transactions on Computational Social Systems},
  year={2025},
  publisher={IEEE}
}

@article{yazdinejad2024hybrid,
  title={Hybrid privacy preserving federated learning against irregular users in next-generation Internet of Things},
  author={Yazdinejad, Abbas and Dehghantanha, Ali and Srivastava, Gautam and Karimipour, Hadis and Parizi, Reza M},
  journal={Journal of Systems Architecture},
  volume={148},
  pages={103088},
  year={2024},
  publisher={Elsevier}
}

@inproceedings{xu2025achieving,
title={Achieving byzantine-resilient federated learning via layer-adaptive sparsified model aggregation},
author={Xu, Jiahao and Zhang, Zikai and Hu, Rui},
booktitle={IEEE Winter Conf. Appl. Comput. Vis. (WACV)},
pages={1508--1517},
year={2025}
}

@inproceedings{xu2022byzantine,
title={Byzantine-robust federated learning through collaborative malicious gradient filtering},
author={Xu, Jian and Huang, Shao-Lun and Song, Linqi and Lan, Tian},
booktitle={IEEE Int. Conf. Distrib. Comput. Syst. (ICDCS)},
pages={1223--1235},
year={2022}
}

@article{blanchard2017machine,
title={Machine learning with adversaries: Byzantine tolerant gradient descent},
author={Blanchard, Peva and El Mhamdi, El Mahdi and Guerraoui, Rachid and Stainer, Julien},
journal={Adv. Neural Inf. Process. Syst. (NeurIPS)},
volume={30},
year={2017}
}

@article{yazdinejad2023ap2fl,
  title={AP2FL: Auditable privacy-preserving federated learning framework for electronics in healthcare},
  author={Yazdinejad, Abbas and Dehghantanha, Ali and Srivastava, Gautam},
  journal={IEEE Transactions on Consumer Electronics},
  volume={70},
  number={1},
  pages={2527--2535},
  year={2023},
  publisher={IEEE}
}

@article{yazdinejad2025breaking,
  title={Breaking Interprovincial Data Silos: How Federated Learning Can Unlock Canada's Public Health Potential},
  author={Yazdinejad, Abbas and Kong, Jude Dzevela},
  journal={Available at SSRN 5247328},
  year={2025}
}

@article{yazdinejad2024robust,
  title={A robust privacy-preserving federated learning model against model poisoning attacks},
  author={Yazdinejad, Abbas and Dehghantanha, Ali and Karimipour, Hadis and Srivastava, Gautam and Parizi, Reza M},
  journal={IEEE Transactions on Information Forensics and Security},
  volume={19},
  pages={6693--6708},
  year={2024},
  publisher={IEEE}
}

@article{yazdinejad2025advanced,
  title={Advanced AI-driven methane emission detection, quantification, and localization in Canada: A hybrid multi-source fusion framework},
  author={Yazdinejad, Abbas and Wang, Hao and Kong, Jude},
  journal={Science of The Total Environment},
  volume={998},
  pages={180142},
  year={2025},
  publisher={Elsevier}
}

@article{yazdinejad2022block,
  title={Block hunter: Federated learning for cyber threat hunting in blockchain-based iiot networks},
  author={Yazdinejad, Abbas and Dehghantanha, Ali and Parizi, Reza M and Hammoudeh, Mohammad and Karimipour, Hadis and Srivastava, Gautam},
  journal={IEEE Transactions on Industrial Informatics},
  volume={18},
  number={11},
  pages={8356--8366},
  year={2022},
  publisher={IEEE}
}

@inproceedings{thanh2020catastrophic,
  title={Catastrophic forgetting and mode collapse in GANs},
  author={Thanh-Tung, Hoang and Tran, Truyen},
  booktitle={2020 international joint conference on neural networks (ijcnn)},
  pages={1--10},
  year={2020},
  organization={IEEE}
}
\bibliographystyle{IEEEtran} 

\vfill

\end{document}